\documentclass[a4paper,11pt,twoside]{article}
\usepackage{hyperref}
\usepackage{amsmath,amssymb,epsf}
\usepackage{amsfonts,amsbsy,amscd,stmaryrd}
\usepackage{paralist}
\usepackage{latexsym,amsopn}
\usepackage{verbatim}
\usepackage{amsthm}
\usepackage{times}
\usepackage{geometry}
\usepackage{setspace}
\usepackage{fancyhdr}

\makeatletter
\renewcommand{\@maketitle}{
\newpage
 \null
 \vskip 1.6em%
 \begin{center}%
  {\LARGE \@title \par}
 \end{center}%
 \vskip 1.4em%
 \begin{center}%
  {\@author \par}%
 \end{center}%
 \par} \makeatother

\newcommand{\half}{\frac{1}{2}}

\newcommand{\bra}{\langle} 
\newcommand{\ket}{\rangle}
\renewcommand\Im{{\rm Im\,}}
\renewcommand\Re{{\rm Re\,}}

\newcommand{\bk}{{\bf k}}

\newcommand{\bx}{{\bf x}}

\newcommand{\e}{{\rm e}}

\renewcommand{\d}{{\rm d}}
\renewcommand{\sp}{{\sigma\,}}
\newcommand{\rs}{{\rm rs\,}}

\def\slim{{\rm s-}\lim}

\def\Ran{{\rm Ran\,}}

\def\retadv{{\rm ret/adv}}

\newcounter{resultcounter}[section]
\renewcommand{\theresultcounter}{\arabic{section}.\arabic{resultcounter}}

\newtheorem{thm}{Theorem}[section]
\newtheorem{theorem}[thm]{Theorem}
\newtheorem{lemma}[thm]{Lemma}
\newtheorem{proposition}[thm]{Proposition}
\newtheorem{assumption}[thm]{Assumption}
\theoremstyle{definition}
\newtheorem{definition}[thm]{Definition}
\theoremstyle{remark}
\newtheorem{remark}[thm]{Remark}
\theoremstyle{definition}
\newtheorem{corollary}[thm]{Corollary}

\numberwithin{equation}{section}

\newcommand{\beq}{\begin{equation}}
\newcommand{\eeq}{\end{equation}}
\newcommand{\beqa}{\begin{eqnarray}}
\newcommand{\eeqa}{\end{eqnarray}}

\def\one{{\mathchoice {\rm 1\mskip-4mu l} {\rm 1\mskip-4mu l} {\rm
      1\mskip-4.5mu l} {\rm 1\mskip-5mu l}}}

\def\cD{{\mathcal D}} \def\cE{{\mathcal E}} \def\cF{{\mathcal F}}
 \def\cH{{\mathcal H}} 
\def\cJ{{\mathcal J}} \def\cK{{\mathcal K}} 
\def\cM{{\mathcal M}}  \def\cO{{\mathcal O}}
  
\def\cS{{\mathcal S}}  
\def\cV{{\mathcal V}}  
 \def\cZ{{\mathcal Z}}
\def\cM{\R^{1,d}}

\newcommand{\R}{{\mathbb R}}
\newcommand{\N}{{\mathbb N}}

\newcommand{\C}{{\mathbb C}}

\def\proof{\noindent{\bf Proof.}\ \ }

\newenvironment{innerlist}[1][\enskip\textbullet]%
        {\begin{compactitem}[#1]}{\end{compactitem}}
        {\begin{compactitem}[#1]}{\end{compactitem}}

\newcommand{\wf}{{\rm WF}}

\newcommand{\supp}{{\rm supp}}
\newcommand{\suppp}{{\rm supp\,}}
\newcommand{\sgn}{{\rm sgn}}

\newcommand{\ccf}{C_{\rm c}^\infty}
\renewcommand{\csc}{C_{\rm sc}^\infty}

\newcommand{\cf}{C^\infty}

        {\begin{compactitem}[#1]}{\end{compactitem}}

\newcommand\cfull{C^{\infty}(\R^{1,d},\mathcal{E})}
\newcommand\dfull{{\mathcal D}(\R^{1,d}\times\R^{1,d},L(\mathcal{E}))'}
\newcommand\dstatic{{\mathcal D}(\R^{1+2d},L(\mathcal{E}))'}

\renewcommand{\i}{{\rm i}}

\begin{document}
\title{Quantum field theory in static external potentials and Hadamard states}
\author{\textsc{Micha\l\ Wrochna}}

\date{}
\maketitle
\begin{center}\small
RTG ``Mathematical Structures in Modern Quantum Physics''\\ Mathematisches Institut, Universit\"at G\"ottingen \\ Bunsenstr. 3-5, D - 37073 G\"ottingen, Germany\\ \vspace{0.1cm} e-mail: \texttt{wrochna@uni-math.gwdg.de} \vspace{0.5cm} \end{center}

\begin{abstract}
We prove that the ground state for the Dirac equation on Minkowski space in static, smooth external potentials satisfies the Hadamard condition. We show that it follows from a condition on the support of the Fourier transform of the corresponding positive frequency solution. Using a Krein space formalism, we establish an analogous result in the Klein-Gordon case for a wide class of smooth potentials. Finally, we investigate overcritical potentials, i.e. which admit no ground states. It turns out, that numerous Hadamard states can be constructed by mimicking the construction of ground states, but this leads to a naturally distinguished one only under more restrictive assumptions on the potentials.
\end{abstract}

\vspace{0.5cm}

\setlength{\parindent}{0cm}{\small {\bf \textit{Acknowledgements}} --- \ It is a pleasure to thank D.~Bahns for many useful remarks and careful reading of the manuscript, as well as J.~Zahn for helpful discussions and valuable comments. The author is grateful to J.~Derezi\'nski and C.~G\'erard for making the manuscript of the book \cite{derger} available before publication. Financial support of the RTG 1493 is gratefully acknowledged.}
\setlength{\parindent}{0.6cm}

\section{Introduction}

\hspace{1cm}It has been realized a long time ago that quantum field theory in the presence of external, classical potentials has much in common with quantum field theory on curved backgrounds at its very foundations. In both theories, the lack of Poincar\'e invariance in the free equations of motion deprives us of a seemingly natural way to specify what should a `vacuum state' be. In some situations, successful and mathematically appealing resolutions of this problem have been found nevertheless and the quantized non-interacting theory has been raised to a fully satisfactory level. 

This includes the Dirac equation on Minkowski space coupled to \emph{static} external potentials (i.e. not depending on time). For a wide range of physically relevant potentials (including arbitrarily strong, smooth ones), it is possible to bring the minimally coupled Dirac equation to the form of an evolution equation governed by a self-adjoint operator on a Hilbert space -- the `minimally coupled Dirac Hamiltonian' $h$. Similar statements are valid for the Dirac equation on a static, globally hyperbolic, smooth manifold \cite{jin}. The construction of the fermionic Fock space and the implementation of the dynamics mimics the procedure used in the absence of potentials \cite{scharf}. In fact, this amounts to performing what one calls \emph{positive energy quantization} of a dynamics and is described in many textbooks and articles with varying degree of generality, see e.g. \cite{positive, baez}. A key ingredient are spectral projections on the positive and negative frequency part of the spectrum of $h$, which enter the construction of the one-particle structure. The outcome includes a \emph{ground state}, which replaces the notion of a `vacuum' and is invariant under the dynamics, implemented as a one-parameter group of unitaries with \emph{positive} generator.

The development of QFT on generic globally-hyperbolic space-times led to a proposal based on a different mathematical setup \cite{wald}. This general quantization program relies on finding bi-solutions (or more generally, bi-parametrices) of the free equations of motion with the same singular structure as the positive-frequency solution on flat Minkowski space. Then, they are used to construct quasi-free states, called Hadamard states, and the requirement on the singularities of the underlying solutions is called the \emph{Hadamard condition}. A breaking point was the observation of M.J.~Radzikowski \cite{radzikowski}, which allowed to rephrase this condition in the language of \emph{microlocal analysis}, using the notion of the \emph{wave front set}. Ultimately, it turned out that this provides the right basis to construct interacting quantum field theories as well \cite{BF00}. The formulation of the Hadamard condition, originally set for scalar fields, has been extended to the spinor case by K.~Kratzert \cite{kratz}, S.~Hollands \cite{hollands}, H.~Sahlmann and R.~Verch \cite{SV01} (see also the comments in \cite{sanders}). It is worth emphasizing that physically, the Hadamard condition (called also `microlocal spectrum condition') is interpreted as an asymptotic positivity condition on the energy and as such it does not fix the state uniquely. In general, there is no distinguished one unless the space-time has specific symmetries or asymptotic symmetries (see \cite{DMP} for asymptotically flat spacetimes).

It is of high importance to know what is the exact relation between the two existing approaches to quantization. More specifically, the following question arises: \emph{Do ground states in non-interacting QFT satisfy the Hadamard condition?} This is true for the Dirac and Klein-Gordon equation on flat Minkowski space under the assumption that no external potentials are present. It has been proved by H.~Sahlmann and R.~Verch \cite{SV00,SV01} that this remains true on stationary space-times. Later, P.~Marecki argued \cite{marecki} that the approach for QFT in generic curved backgrounds can be fully adapted to QFT in external potentials and initiated such a program, but left the question of Hadamard condition for ground states unsolved. The main aim of this paper is precisely to solve this problem. We prove that ground states for the Dirac equation on Minkowski space with static, smooth potentials do satisfy the Hadamard condition, as conjectured by Marecki.\\

For the Klein-Gordon equation, the situation is yet more complicated. The problem is similar to the Dirac equation case, except that in order to get a ground state, restricting to static potentials and stationary metric is not sufficient. Ground states do not exist if the electric potential exceeds some critical value, or if the metric is `superradiant', see for instance the appendix in \cite{fulling} for a detailed description (and \cite{kaywald} specifically for superradiant black-hole spacetimes). Phenomena characteristic for those situations are altogether named \emph{Klein paradox} and stem from the impossibility of describing the classical dynamics as a unitary group. In this paper, we are interested in both the subcritical and overcritical cases. We show that ground states for the Klein-Gordon equation on Minkowski space with subcritical, static, smooth potentials satisfy the Hadamard condition. For overcritical potentials, we still find families of Hadamard states and investigate if there is a natural way of distinguishing one of them. The main difference between the states obtained in the subcritical and the overcritical case is that the latter are not faithful ones. \\

Before displaying our results in more detail, let us first point out the main motivations behind attempts at merging techniques from QFT on curved backgrounds and QFT in external potentials. We would like to stress that in each of the possible applications we mention, existence of distinguished Hadamard states is a question of critical importance and our result for static potentials provides a first answer to it.\\
\begin{itemize}
\item Adaptation of \textit{causal renormalization} to QFT in curved space-time has been an astonishing success of perturbative theory \cite{BF00}. The powerful tools used there can be adapted to QFT in external potentials (see \cite{dosch} for renormalization of QED in static external fields) and in particular allow for a local definition of Wick and time-ordered products \cite{marecki}. It is desirable to extend this formalism to more general cases and make the connection with commonly used techniques in QFT in strong external potentials, such as Furry picture QED \cite{fradkin,mohr}, and explain to what extent they can be mathematically well-posed. 

\item Renormalizability of field theories on non-commutative Moyal space is another related problem, where external potentials and microlocal properties of associated distributions come into play. In the Euclidean formalism, a  successful renormalization program has been established for specific models involving a potential term, and their Minkowskian analogues are thus natural candidates for renormalizable theories. A direct transition from the Euclidean to the Minkowski signature by a Wick rotation has been shown to be problematic for vanishing potentials \cite{bahns} and the general case is unknown. Consequently, one is interested in the question whether the Minkowskian versions of the models with potential terms can be consistently treated by extending the causal renormalization machinery, not referring to Euclidean techniques in the process. Unfortunately, defining a meaningful quantum field theory in an external potential, such as the one suggested by the Grosse--Wulkenhaar model, poses difficulties appearing already at the non-interacting level, due to the `superradiant' nature of the potential and its dependence on time. Even disregarding the peculiarities of the non-interacting theory, computations in perturbative theory are plagued by divergences, which can be traced back to analytic properties of the solutions of the Klein-Gordon equation with external potential \cite{zahn}. Therefore, a systematic study of the latter is needed.

\item Ultimately, QFT in external potentials may prove to be an important guide in understanding \textit{back-reaction effects} of quantum fields on curved backgrounds. The existing proposal for describing back-reaction on curved backgrounds, formulated by R.~Wald in \cite{wald77}, is based on the so-called \emph{semi-classical Einstein equation}, involving the renormalized quantum stress-energy tensor \,:$T_{\mu\nu}(x)$: (see \cite{moretti} and \cite{hw05} for an up to date discussion) evaluated on a suitable state and describing its influence on the background metric. This concept can be modified to describe back-reaction of quantum fields on external potentials, by looking for solutions of \emph{semi-classical Maxwell equations}, involving the renormalized quantum current operator \,:$j^{\mu}(x)$: \cite{marecki} and describing its influence on the background electromagnetic field. Such a back-reaction theory should approximate the fully quantized theory, specifically spinor or scalar QED, with a better accuracy than QFT in fixed external backgrounds. The verification of this claim can provide useful indication for the back-reaction theory based on the semi-classical Einstein equation, which is believed to be a first approximation of quantum gravity.

Although it seems difficult to compare such a semi-classical QED with its fully quantized counterpart and even with other back-reaction approximations such as the mean field method \cite{meanfield}, decisive answers could be provided by experimental data. Electromagnetic fields strong enough to carry sizable back-reaction effects are expected to be produced in a laboratory setup in the near future \cite{ringwald, ruffini}.
\end{itemize}

The content of this paper can be summed up as follow. 

In Section \ref{section2}, we start by introducing the basic definitions and notations relevant for the flat static case. Then, we show that the Hadamard condition in the flat, static case is implied by an asymptotic condition on the support of the Fourier transform of the candidates for positive and negative frequency solutions (\emph{Theorem \ref{basetheorem}}). The proof relies on well-known or elementary facts on the wave front set and no pseudo-differential operator techniques are directly employed.

In Section \ref{section3}, we consider the Dirac case and rephrase the usual construction of ground states in terms of positive and negative frequency solutions $S_+,S_-$. We show that they satisfy the condition from Section \ref{section2} and positivity (\emph{Proposition \ref{mainproposition}}). The support property of the Fourier transforms of $S_+$, $S_-$ is part of the folklore knowledge and is easy to show, but in view of Section \ref{section2} this suffices us to prove Marecki's conjecture, i.e. that $S_+$ satisfies the Hadamard condition.

Section \ref{section4} deals with the Klein-Gordon equation. Motivated by the results of Section \ref{section3}, we follow as closely as possible the construction used for the Dirac case. The role of the analogue of the minimally coupled Dirac Hamiltonian is then played by a \textit{Krein self-adjoint} operator $b$ in a \textit{Krein space} (called also `Hilbert space with indefinite inner product' in the physics literature). We use extensively  results due to H.~Langer, B.~Najman and C.~Tretter \cite{LNT} and C.~G\'erard \cite{gerard} for the operator $b$ to construct ground states in the subcritical case. This construction differs from what is usually presented in the literature and although it leads to the same result, it has the advantage of being extendible to the overcritical case with almost no change. In the overcritical case, the outcome is not a ground state, but a family of Hadamard states instead. We show that under additional restrictions on the potentials, there is a natural way of specifying a distinguished Hadamard state (\emph{Corollary \ref{finalcoro}}). We then discuss second quantization using the obtained Hadamard states. An outlook is presented in Section \ref{section5}.

In Appendix \ref{s:appendixa}, the definition of the wave front set and its basic properties are briefly recalled. 

\subsection{Notations}

$\cE$ will always denote a finite dimensional vector space and $L(\cE)$ the space of its endomorphisms. The space of $\cE$-valued test functions in $\R^p$ is denoted $\ccf(\R^p,\cE)$ and its dual $\cD(\R^p,\cE)'$ or $\cD'(\R^p)$ if $\cE=\C$. The space of Schwartz functions is denoted $\cS(\R^p,\cE)$, its dual $\cS(\R^p,\cE)'$. We often use a function-like notation $u(x)$ to denote a distribution $u\in\cD(\R^p,\cE)'$, whereas $\bra u, \varphi \ket$ (or alternatively $\int_{\R^p} u(x)\varphi(x)\d x$) is used to denote the action of $u$ on a test function $\varphi$. The support of $u\in\cD(\R^p,\cE)'$ is denoted $\suppp u$.

We denote $\cM$ the Minkowski space with $d\geq1$ spatial dimensions, signature $(-,+,\dots,+)$, and set $n:=1+d$. The causal future (resp. causal past) of $K\subset\cM$ is denoted $J^+(K)$ (resp. $J^-(K)$). The space of smooth spacelike compact functions is by definition
\[
\cf_{\rm sc}(\cM,\cE):=\{ f\in\cf(\cM,\cE) : \  \exists K \subset \cM \mbox{\ compact\ s.t.\ }\ \supp f \subset J_+(K)\cup J_-(K) \}.
\]

Unless specified otherwise, we always consider differential operators with \textit{smooth} coefficients. Often, we write $P_{(x)}$ for a differential operator acting in the variables $x$.

For a Hilbert space $\cH$, we denote $B(\cH)$ the set of bounded operators on $\cH$ and $1\in B(\cH)$ the identity operator. Given a linear operator $a$ in $\cH$, $\cD(a)$ denotes its domain, $\Ran a$ its range, $\sigma(a)$ its spectrum, $\sigma_{\rm p}(a)$ its point spectrum and $\sigma_{\rm ess}(a)$ its essential spectrum.

$\cJ\subset \R$ will always denote a finite union of intervals. We denote $\overline{\cJ}$ its closure, $\partial\cJ$ its boundary and $\one_{\cJ}$ its characteristic function.

\section{Hadamard condition in the static case}\label{section2}

\setlength{\parindent}{0cm}
\setlength{\parskip}{6pt plus 2pt minus 1pt}

\subsection{Causal propagator}

Let $\cE$ be a finite dimensional vector space. We introduce basic notions for hyperbolic differential operators on $\cM$ with time-independent coefficients. 

\begin{definition}
Let $D:\cf(\cM,\cE)\to\cf(\cM,\cE)$ be a differential operator and denote $D^*$ its formal adjoint. Distributions $\tilde{S}_{\retadv}\in\cD(\cM\times\cM,L(\cE))'$ are called \emph{retarded} / resp. \emph{advanced fundamental solutions} for $D$, if they satisfy
\begin{eqnarray*}
{\rm (1)} &  D_{(x)}\tilde{S}_{\retadv}(x,x') = \delta(x,x')\cdot 1, \\
{\rm (2)} &  D^*_{(x')}\tilde{S}_{\retadv}(x,x') = \delta(x,x') \cdot 1,\\
{\rm (3)} & \suppp \tilde{S}_{\retadv} \subset \{ (x,x') : \ x\in J^{+ / -}(x')\},
\end{eqnarray*}
where $1$ is the identity in $L(\cE)$.
\end{definition}

\begin{definition} A differential operator $P:\cf(\cM,\cE)\to\cf(\cM,\cE)$ is said to be \emph{normally hyperbolic} of order two in $\cM$, if it is of the form $\partial_{\mu}\partial^{\mu}+K^{\mu}(x)\partial_{\mu}+L(x)$ for some $K_{\mu},L\in\cf(\cM,\cE)$, $\mu=1,\dots,n$.
\end{definition}

\begin{definition}\label{defprenormally}We say that a hyper\-bolic dif\-ferential ope\-rator $D:\cf(\cM,\cE)\to\cf(\cM,\cE)$ is \emph{prenormally hyperbolic} if there exists a hyperbolic differential operator $D'$ such that both $D D'$ and $D'D$ are normally hyperbolic of order two.
\end{definition}

By convention, we allow $D'$ to be of order zero, so that normally hyperbolic operators of order two are prenormally hyperbolic in the sense above. Note that this differs from the definition proposed in \cite{muehlhoff}. If $D$ is prenormally hyperbolic, one can use the existence and uniqueness theorem of retarded/advanced fundamental solutions available for normally hyperbolic operators $D D'$ and $D'D$ to construct unique retarded/advanced fundamental solutions for $D$. This has been performed by J.~Dimock for the case of the Dirac equation in globally-hyperbolic space-times \cite{dimock} and generalized by R.~M\"uhlhoff \cite{muehlhoff}.

\begin{theorem}\label{theorem:spropagators}Let $D$ be a prenormally hyperbolic differential operator. Then, $\tilde{S}_{\retadv}(x,x'):=D'_{(x)}\tilde\Delta_{\retadv}(x,x')\in\dfull$ is the unique retarded/advanced fundamental solution for $D$, where $\tilde\Delta_{\retadv}\in\dfull$ is the retarded/advanced fundamental solution for the normally hyperbolic operator of order two $DD'$.
\end{theorem}
\proof It suffices to follow step by step the proof of \cite[Theorem 1]{muehlhoff}. It is assumed there that $D$ is of order $1$, but this is in fact not necessary.
\qed

The \emph{causal propagator} $\tilde{S}$ is defined as $\tilde{S}:=\tilde{S}_{\rm ret}-\tilde{S}_{\rm adv}$.

The theorem below, characterizing the wave front set of $\tilde{S}$, is a straightforward consequence of Theorem \ref{theoremchar} and \cite[Theorem 4.16]{BF09}. For the definition of the wave front set and its properties, see Appendix A.  

\begin{theorem} Let $\tilde{S}\in\dfull$ be the causal propagator for a prenormally hyperbolic operator $D:\cfull\to\cfull$. Then,
\begin{eqnarray*}
\wf(\tilde{S})= \{ (x,x',k,-k): \ x\neq x', \ x-x' \mbox{\,and\ } k \mbox{\ lightlike}, k \mbox{\ coparallel\ to\ } x-x'  \}
\\ \cup \{ (x,x,k,-k): \ k \mbox{\ lightlike}, k\neq 0 \}.
\end{eqnarray*}
\end{theorem}

In this paper, we are interested in the case when $D_{(x)}=D_{(t,\bx)}$ is a differential operator with coefficients not depending on $t$. Then, $\tilde{S}(t,\bx,t',\bx')\in\cD(\R^{1,d}\times\R^{1,d},L(\cE))$ depends on the time coordinates via the difference $t-t'$ only. More precisely, $\tilde{S}(t,\bx,t',\bx')$ is the pullback of a distribution in $\cD(\R^{1+2d},L(\cE))'$ under the map 
\beq\label{maptau}
\R^{1,d}\times\R^{1,d}\ni(t,\bx,t',\bx')\mapsto \tau(t,\bx,t',\bx'):=(t-t',\bx,\bx')\in \R^{1+d+d}
\eeq
and we denote this distribution $S(t,\bx,\bx')$. We also call it the \emph{causal propagator}. Its wave front set is given by the relation
\[
\wf(\tilde{S})=\{ (t,\bx,t',\bx';\xi,\bk,-\xi,\bk') :  \ (t-t',\bx,\bx';\xi,\bk,\bk') \in \wf(S) \}.
\]

 Going further with a time-translation invariant convention, we say that a distribution $u\in\cD(\R^{1+2d},L(\cE))'$ is a \emph{bi-parametrix} for $D_{(x)}$, if 
\[
D_{(t,\bx)}u(t-t',\bx,\bx')=f(t,t',\bx,\bx'), \ \ \ D_{(t',\bx')}u(t-t',\bx,\bx')=g(t,t',\bx,\bx').
\]
for some $f,g\in\cf(\cM,\cE)$, and a \emph{bi-solution} if $f=g=0$.

The frequency part of the wave front set of a solution of a normally hyperbolic differential operator is contained in the light cone (as follows from Theorem \ref{theoremchar}), which translated to a bi-solution $u$ for $D$ means in our convention that
\beq
(t,\bx,\bx';\xi,\bk,\bk')\in \wf(u) \Rightarrow \xi^2-\bk^2=\xi^2-\bk'^2=0. \label{eq:eliminate}
\eeq
Let us remark that this excludes points with $\xi=0$ from the wave front set, since it would imply $\bk^2=\bk'^2=0$ and points of the form $(t,\bx,\bx';0,0,0)$ are not in $\wf(u)$ by definition. 

\subsection{Hadamard condition in the static case}

Assume $D:\cf(\cM,\cE)\to\cf(\cM,\cE)$ is a prenormally hyperbolic differential operator with coefficients not depending on time. Let $S\in\dstatic$ be the causal propagator for $D$ and $S_+$ a bi-parametrix for $D$.

In the static case, employing the notations introduced before, the Hadamard condition can be written as follows:

\begin{definition}$S_+\in\dstatic$ is said to satisfy the \emph{Hadamard condition} if 
\[
\wf(S_+)=\wf(S)\cap K^+,
\]
where $K^+:=\R^{1+2d}\times (0,\infty)\times \R^{2d}$.
\end{definition}
We often call bi-solutions satisfying the Hadamard condition \emph{positive-frequency solutions}. Of particular interest are bi-solutions or bi-parametrices satisfying additionally a positivity condition, which ensures that they determine unambiguously a quasi-free state (called \emph{Hadamard state}) on an adequate $^*$-algebra of CCR/CAR relations, as explained for instance in \cite{BF09}. Throughout this paper, we concentrate on such bi-solutions rather than on the corresponding states.

A key point in our investigations is the observation, that the static Hadamard condition is implied by a stronger condition on the support of the Fourier transform of $S_+$ and $S_-$. This will turn out to be easily proved to hold in our case of interest.

We will consider distributions contained in the topological tensor product $\cS'(\R)\hat\otimes\cD(\R^{2d},L(\cE))'$ $\subset$ $\cD(\R^{1+2d},L(\cE))'$, so that it makes sense to apply a Fourier transform in the first argument (thanks to continuity of the Fourier transform in $\cS'(\R)$).

\begin{definition}\label{defspectral}Let $S\in\dstatic$ be the causal propagator for $D$ and let $S_+\in\cS'(\R)\hat\otimes\cD(\R^{2d},L(\cE))'$ be a bi-parametrix for $D$. We say that $S_+$ satisfies the \emph{static asymptotic spectral condition}, if
\beq
S=S_+ + S_- + S_0,
\eeq
where $S_0\in\cf(\R^{1+2d},L(\cE))$, $S_-\in \cS'(\R)\hat\otimes\cD(\R^{2d},L(\cE))'$ is a bi-parametrix for $D$ and
\begin{enumerate}\setlength{\itemsep}{-1.5pt}
\item $\supp(\cF_0 S_+)\subset [\alpha, \infty)\times\R^{2d}$ for some $\alpha\in\R$,
\item $\supp(\cF_0 S_-)\subset (-\infty,\beta]\times\R^{2d}$ for some $\beta\in\R$,
\end{enumerate}
where $\cF_0$ is the Fourier transform in the first variable.
\end{definition}

Note that we do not impose $\alpha\geq \beta$. Therefore, the situation described by this condition can be understood as a splitting of $S$ into a positive and negative frequency part in an asymptotic sense, i.e. accurate for sufficiently high frequencies. 

\begin{theorem}\label{basetheorem}Let $D:\cfull\to\cfull$ be a prenormally hyperbolic differential operator (in the sense of Definition \ref{defprenormally}) with coefficients not depending on time and let $S_+\in\dstatic$ be a bi-parametrix for $D$. Then, the static asymptotic spectral condition implies the Hadamard condition.
\end{theorem}

In the proof, we make use of the following lemma due to L.~H\"ormander.

\begin{lemma}[\cite{hoermander}, Lemma 8.1.7]\label{limitcone}
If $u\in\cS'(\R^p)$, then $\wf(u)\subset\R^p\times F$, where 
\[
F:=\left\{ \lim_{j\to\infty} \alpha_j x_j : \ x_j\in\supp( \cF u)\subset\R^p, \ \alpha_j>0, \ \lim_{j\to\infty}\alpha_j=0\right\}
\]
and $\cF$ denotes the Fourier transform in $\R^p$. 
\end{lemma}

As a corollary, we get:

\begin{lemma}\label{argument} Let $u\in\cS'(\R)\hat{\otimes}\cD(\R^{q},\cE)'$ and denote $\cF_0 u$ its Fourier transform in the first argument. Then,
\begin{eqnarray}
\supp(\cF_0 u)\subset [\alpha, \infty)\times\R^q \mbox{\ for\ some\ } \alpha\in\R \ \Rightarrow \ \wf(u)\subset K^+\cup K^0,\label{argument1}\\
\supp(\cF_0 u)\subset (-\infty,\beta]\times\R^q \mbox{\ for\ some\ }\beta\in\R \ \Rightarrow \ \wf(u)\subset K^-\cup K^0,\label{argument2}
\end{eqnarray}
where $K^{\pm}=\R^{1+q}\times((0,\pm\infty)\times\R^q)$, $K^{0}:=\R^{1+q}\times(\{0\}\times\R^q)$.
\end{lemma}
\proof
Assume $u\in\cS(\R^{1+q},\cE)'$. We can also assume without loss of generality $\cE=\C$. Then $\supp(\cF_0 u)\subset [\alpha, \infty)\times\R^q$ implies $\supp(\cF u)\subset [\alpha, \infty)\times\R^q$, where $\cF$ denotes the Fourier transform in all the $1+q$ variables. Claim (\ref{argument1}) (and analogously (\ref{argument2})) follows directly from Lemma \ref{limitcone}.

The general case $u\in\cS'(\R)\hat{\otimes}\cD(\R^{q},\cE)'$ follows by applying the preceding arguments to $(\one\otimes\chi)u\in\cS(\R^{1+q},\cE)'$ for each $\chi\in\ccf(\R^q)$.
\qed

\textbf{Proof of Theorem \ref{basetheorem}.} \ Assume $S_+$ satisfies the static asymptotic spectral condition. Lemma \ref{argument} implies directly
\beq\label{proofeq1}
\wf(S_{\pm})\subset K^{\pm}\cup K^0.
\eeq
Because $S_{+}$ and $S_-$ are bi-solutions for $D$, $\wf(S_{\pm})\cap K^0=\emptyset$, as can be seen from (\ref{eq:eliminate}). Consequently, (\ref{proofeq1}) is equivalent to $\wf(S_{\pm})\subset K^{\pm}$. In particular, $\wf(S_+)$ and $\wf(S_-)$ are disjoint. Together with $S-S_0=S_++S_-$, this entails precisely
\[
\wf(S_{\pm})=\wf(S-S_0)\cap K^{\pm}=\wf(S)\cap K^{\pm}.
\]
\qed

Let us point out the following conclusion from the proof of Theorem \ref{basetheorem}: given $S$ and a candidate for a positive-frequency solution $S_+$, it suffices to prove (\ref{proofeq1}) to check that the Hadamard condition is satisfied. It is a considerable simplification, as the exact form of the whole wave front set of $S$ and $S_+$ is not important in doing so. Such kind of argument can be rewritten for the non-static and curved case as well, a similar idea was in fact used in \cite{reeh} in the context of wave front sets of Hilbert-space valued distributions.

\section{Spin-1/2 case}\label{section3}
The Dirac equation in $d$ spatial dimensions, static external potentials $A_{\mu}(\bx)=(V(\bx),A_{i}(\bx))$ ($i=1,\dots,d$) and variable mass $m=m(\bx)$ is given by:
\[
(\i\partial_{t}+h_{(\bx)})\psi(t,\bx)=0,
\]
where $h_{(\bx)}:\cf(\cM,\C^r)\to\cf(\cM,\C^r)$ is the differential operator
\[
h_{(\bx)}=-\i\sum_{i=1}^d\alpha^i(\partial_{i}-\i A_{i}(\bx))-V(\bx)-m(\bx)\beta,
\]
and $\alpha^i$, $\beta$ are the $r\times r$ Dirac matrices with $r = 2^l$, $l$ being the greatest integer that is not
greater than $n/2$. We assume $A_i(\bx)$, $V(\bx)$, $m(\bx)$ are real valued \emph{smooth functions}.

A well-known theorem states, that the Cauchy problem associated to the Dirac equation is solved by the causal propagator \cite{dimock}.
\begin{theorem}
\label{diraccauchy}Let $\vartheta\in\ccf(\R^d,\C^r)$. There exists a unique $\psi\in\csc(\R^{1,d},\C^r)$ that solves
\[
\begin{cases}(\i\partial_{t}+h_{(\bx)})\psi(t,\bx)=0,\\
\psi(0,\bx)=\vartheta(\bx).
\end{cases}
\]
It is given by
\[
\psi(t,\bx)=\int_{\R^d}S(t,\bx,\bx')\gamma^0\vartheta(\bx')\d \bx',
\]
where $\gamma^0:=-\i \beta$ and $S\in\dstatic$ is the causal propagator for $(\i\partial_t+h_{(\bx)})$.
\end{theorem}

\subsection{Dirac Hamiltonian}

The standard approach to quantizing the non-interacting Dirac field in static external potentials relies on the possibility of assigning to $h_{(\bx)}$ a self-adjoint extension $h$, called the Dirac Hamiltonian, in the Hilbert space $L^2(\R^d,\C^r)$. A treatment for basic classes of potentials, including a description of the domain of $h$ and its spectral properties, can be found for instance in \cite{thaller} for the physically most important case $d=3$. In this paper we are interested in the case of smooth potentials exclusively and under such assumption, as pointed out for instance in \cite{shi}, essential self-adjointness of $h_{(\bx)}$ follows directly from the general arguments given in \cite{chernoff}. Let us stress that no decay at infinity of the potentials is required.

\begin{theorem}If $A_i,V,m\in\cf(\R^d,\R)$, then the operator $h_{(\bx)}$ acting on $\ccf(\R^d,\C^r)$ is essentially self-adjoint in the Hilbert space $L^2(\R^d,\C^r)$.
\end{theorem}
We denote $h$ the closure of $h_{(\bx)}$. 

In particular, the differential expression $h^0_{(\bx)}$, corresponding by definition to $V=A_i=0$ and $m(\bx)\equiv m$, is essentially self-adjoint on $C_{\rm c}^{\infty}(\R^d,\C^r)$. Its closure (the \textit{free Dirac Hamiltonian}), denoted $h^0$, has domain $\cD(h^0)=H^1(\R^d,\C^r)$ and spectrum $\sp (h^0) =(-\infty,-m]\cup[m,\infty)$.

\subsection{Positive and negative frequency solutions}

The family of unitaries $\e^{\i t h}$ solves uniquely the Cauchy problem for the Dirac equation, and hence is directly related to the causal propagator, according to Theorem \ref{diraccauchy}. Formally, the causal propagator $S(t,\bx,\bx')$ is for fixed $t$ the integral kernel of $-\gamma^0\e^{\i t h}$. Propagation of positive-frequency solutions is described by the family $-\gamma^0\e^{\i t h}\one_+(h)$, where $\one_+(h)$ is the projection on the positive part of the spectrum of $h$. The integral kernel $S_+(t,\bx,\bx')$ of $\e^{\i t h}\one_+(h)$ is what one calls the positive-frequency solution $S_+$. We make these statements precise and prove, that $S_+$ defined this way satisfies the static asymptotic spectral condition.

We proceed by defining the multi-linear functionals
\beq\label{eq:defs}
\bra S, f\otimes \bar{u} \otimes v \ket :=-( u | \gamma^0 (\cF^{-1}f)(h) v ), \ \ \ \ \ \bra S_{+},f\otimes \bar{u} \otimes v\ket:=-( u | \gamma^0 (\cF^{-1}f)(h)\one_{+}(h) v )
\eeq
for $f\in\cS(\R)$, $u$, $v\in\cS(\R^d,\C^{r})$. Here, $\one_{+}( \cdot ):=\one_{[0,\infty)}( \cdot )$ is the characteristic function of the closed half-line, $\cF^{-1} f$ is the inverse Fourier transform of $f$ and $(\cF^{-1}f)(h)$, $\one_{+}(h)$ are defined by function calculus. The complex conjugate $\bar{u}$ of $u$ is needed for linearity.

\begin{proposition}\label{mainproposition}$S$ and $S_{+}$ are well-defined distributions in $\cS(\R^{1+2d},L(\C^{r}))'$. Furthermore,
\begin{enumerate}\setlength{\itemsep}{0pt}
\item $S$ and $S_+$ are bi-solutions for $(\i\partial_t+h_{(\bx)})$, i.e
\beqa
(\i\partial_t+h_{(\bx)})S(t,\bx,\bx')=0, \ \ \ \ \ (\i\partial_t+h_{(\bx)})S_+(t,\bx,\bx')=0,\label{ssolution1}\\
(-\i\partial_t+h_{(\bx')})S(t,\bx,\bx')=0, \ \ \ \ \ (-\i\partial_t+h_{(\bx')})S_+(t,\bx,\bx')=0\label{ssolution2}.
\eeqa
\item $S$ is the causal propagator for $(\i\partial_t+h_{(\bx)})$,
\item $S_+$ satisfies the static asymptotic spectral condition. More precisely
\begin{eqnarray}\label{toprove3}
\supp (\cF_0 S_{\pm})\subset [0,\pm\infty)\times\R^{d+d},
\end{eqnarray}
where $S_-:=S-S_+$.
\item $S_+$ satisfies the following \emph{positivity condition}
\beq\label{spositivity}
-(\tau^* S_+)(\overline{F} \otimes \gamma^0 F)\geq 0 \ \ \ \forall F\in\ccf(\R^{1,d},\C^r),
\eeq
where $\tau$ is the map defined by (\ref{maptau}).
\end{enumerate}

\end{proposition}
\proof
By the Schwartz kernel theorem, (\ref{eq:defs}) defines uniquely a tempered distribution if $S_+:\cS(\R)\otimes\cS(\R^d,\C^{r})\otimes\cS(\R^d,\C^{r})\to\C$ is continuous. By Schwarz inequality and function calculus of self-adjoint operators (for Borel functions), we have
\begin{eqnarray*}
|\bra S_+, f \otimes \bar{u} \otimes v \ket |=|( u | \gamma^0 (\cF^{-1}f)\one_{+}(h) v )|\leq \| u\| \|\gamma^0(\cF^{-1}f)(h)\one_{+}(h)\| \left\|v\right\| \\ \leq \left\| u\right\| \|\gamma^0(\cF^{-1}f)\one_{+}\|_{\infty} \left\|v\right\| \leq\left\|  u\right\| \|\gamma^0\cF^{-1}f\|_{\infty}\left\|v\right\|.
\end{eqnarray*}
Convergence of $f$ to $0$ in $\cS(\R)$ implies $ \cF^{-1} f\to0$ in $\cS(\R)$ and consequently $\|  \cF^{-1} f \|_{\infty}\to 0$. Furthermore, convergence of $u$ (resp. $v$) to $0$ in $\cS(\R^d,\C^r)$ implies $\left\|u\right\|\to 0$ (resp. $\|v\|\to 0$), hence $S_+$ is continuous. The reasoning for $S$ is analogous.
\begin{enumerate}
\item To prove (\ref{ssolution1}) it suffices to check the equality on simple tensors. For arbitrary $f\in\ccf(\R)$, $u$, $v\in\ccf(\R^d,\C^r)$, we have 
\[
\bra h_{(\bx)}S, f\otimes \bar{u} \otimes v\ket =-(u|\gamma^0 (\cF^{-1}f)(h)h v),
\]
\[
\bra \partial_t S, f\otimes \bar{u} \otimes v\ket = -(u|\gamma^0 (\cF^{-1}\partial_t f)(h)v)=\i (u| \gamma^0 (\cF^{-1}f)(h) hv),
\]
where we used $v\in\ccf(\R^d,\C^r)\subset\cD(h)$. The remaining assertions follow in an analogous way.

\item This is a straightforward consequence of Theorem \ref{diraccauchy}, as $S$ solves the same Cauchy problem as the causal propagator.

\item For any $\varphi\in\ccf(\R)$ with $\supp\varphi\subset(0,-\infty)$, we have
\[
\bra \cF_0 S_, \varphi \otimes \bar{u} \otimes v \ket =-( u | \gamma^0 \varphi(h)\one_{+}(h) v )=-( u | \gamma^0 (\varphi\cdot\one_{+})(h) v )=0
\]
and analogously for $S_-$, hence (\ref{toprove3}).
\item For each $f\in\ccf(\R)$, $u\in\ccf(\R^d,\C^r)$,
\[
\bra -\gamma^0 \tau^* S_+, \overline{f\otimes u} \otimes f\otimes u \ket =( u | ( \overline{\cF^{-1}f}\cF^{-1}f)(h)\one_{+}(h) u )\geq 0.
\]
\end{enumerate}
\qed

\begin{corollary} The distribution $S_+\in\dstatic$ satisfies the Hadamard condition.
\end{corollary}

The construction of the Hadamard state associated to $S_+$, the role of the positivity condition (\ref{spositivity}) and second quantization are described in detail in \cite{hack}, see also \cite{bratteli} for a general overview on quasi-free states.

\section{Spin-0 case}\label{section4}

\subsection{Two-component Klein-Gordon equation}

Consider the Klein-Gordon equation with static potentials and variable mass
\beq\label{kg}
\left[\left(\partial_{t}-\i V(\bx)\right)^2-\sum_{i=1}^d\left(\partial_{i}-\i A_{i}(\bx)\right)^2+m(\bx)^2\right]\phi(t,\bx)=0.
\eeq
We will write formally $\epsilon^2_{(\bx)}:=-\sum_{i=1}^d\left(\partial_{i}-\i A_{i}(\bx)\right)^2+m(\bx)^2$ 
as a differential operator acting on $\ccf(\R^d)$ (and do not assign any meaning to $\epsilon$ for a moment). 

It can be rewritten in a Hamilton form as
\begin{equation}\label{kg2}
 (\i\partial_t+b_{(\bx)}) \begin{pmatrix} v_1(t,\bx) \\ v_2(t,\bx) \end{pmatrix}= \begin{pmatrix} v_1(t,\bx) \\ v_2(t,\bx) \end{pmatrix}, \ \ \ b_{(\bx)}:=\begin{pmatrix} V(\bx) & 1 \\ \epsilon_{(\bx)}^2 & V(\bx)\end{pmatrix}.
\end{equation}

The one-component and two-component Klein-Gordon equations are related as follows: 
\beq\label{onevstwo}
(\i\partial_t+b_{(\bx)})\begin{pmatrix} v_1(t,\bx) \\ v_2(t,\bx) \end{pmatrix}=0 \ \Longleftrightarrow  \ \begin{cases} \left[(\partial_t-\i V(\bx))^2+\epsilon_{(\bx)}^2\right]\phi(t,\bx)=0, \\ v_1=\phi, \ \ \ \ v_2(t,\bx)=-(\i\partial_t+V(\bx))\phi(t,\bx). \end{cases}
\eeq

The differential operator $(\i\partial_t+b_{(\bx)})$ is prenormally hyperbolic (in the sense of Definition \ref{defprenormally}). Indeed, defining
\[
b'_{(\bx)}:=\begin{pmatrix} -V(\bx) & 1 \\ \epsilon_{(\bx)}^2 & -V(\bx)\end{pmatrix},
\]
\[
(-\i\partial_t+b')(\i\partial_t+b)=(\i\partial_t+b)(-\i\partial_t+b')=\begin{pmatrix} (\partial_t-\i V)^2+\epsilon^2 & 0 \\ 0  & (\partial_t-\i V)^2+\epsilon^2 \end{pmatrix}.\]

An analogue of Theorem \ref{diraccauchy} is available, where $\gamma^0$ gets now replaced by $\sigma^0:=\i\begin{pmatrix} 0 & 1 \\ 1 & 0 \end{pmatrix}$.

\begin{theorem}\label{kgcauchy}Let $\vartheta\in\ccf(\R^d,\C^2)$. There exists a unique $\phi\in\csc(\R^{1,d},\C^2)$ that solves
\[
\begin{cases}(\i\partial_t+b_{(\bx)})\phi(t,\bx)=0,\\
\phi(0,\bx)=\vartheta(\bx).
\end{cases}
\]
It is given by
\[
\phi(t,\bx)=\int_{\R^d}S(t,\bx,\bx')\sigma^0\vartheta(\bx')\d \bx',
\]
where $S\in\cD(\R^{1+2d},L(\C^2))'$ is the causal propagator for $\i\partial_t+b_{(\bx)}$.
\end{theorem}

\subsection{Klein-Gordon Hamiltonian --- Introduction}

We proceed with putting the Klein-Gordon equation in a functional analysis framework. To this end, we need to define $\epsilon$ as an operator in the Hilbert space $L^2:=L^2(\R^d)$. Throughout this section, we denote the scalar product in $L^2$ by $( \cdot | \cdot)$. We make the following assumption:

\begin{assumption}\label{mainassumption}The operator $\epsilon^2_{(\bx)}=-\sum_{i=1}^d\left(\partial_{i}-\i A_{i}(\bx)\right)^2+m(\bx)^2$ on $\ccf(\R^d)$ admits a self-adjoint extension denoted $\epsilon^2$, s.t. $\epsilon^2\geq \mu^2 \cdot 1$ for some $\mu>0$.
\end{assumption}

For non-vanishing $A_i$, the operator $\epsilon$ is called a \textit{magnetic Schr\"{o}dinger operator}. An extensive study of magnetic Schr\"{o}dinger operators can be found in \cite{AHS}, criterions for Assumption \ref{mainassumption} are discussed in \cite{shubin} and refer implicitely to the behaviour of $A_i$ at infinity. Later on we will be interested only in potentials satisfying some additional conditions and we postpone the discussion of examples until then.

Note that for (let's say) vanishing $A_i$, Assumption \ref{mainassumption} excludes the massless case $m=0$. The massless case shows indeed peculiarities, which are already visible in the Dirac case. As we will be mainly interested in problems of a whole different kind, caused by the presence of the electric potential, we do not discuss such massless problems and restrict ourselves to referring the interested reader to the work of P.~Jonas \cite{jonas}, in which technical results for Klein-Gordon type equations displaying both kind of difficulties are described.

One would wish to associate to $b_{(\bx)}$ a self-adjoint operator in a Hilbert space in analogy to the Dirac Hamiltonian $h$. This turns out not to be possible. There is a natural symmetric form $[\cdot | \cdot ]$ for which $b$ is formally symmetric, i.e. $[\cdot | b \cdot ]=[b \cdot | \cdot ]$ on suitable elements. It is given by
\beq\label{chargeproduct}
[u|v]:=(u_1|v_2)+(u_2|v_1)
\eeq
for $u=(u_1,u_2),v=(v_1,v_2)\in L^2 \oplus L^2$. In the literature, it is sometimes called the \textit{charge inner product}. Clearly, it is not positive definite and cannot be used directly to define a Hilbert space. One has to refer to Krein spaces techniques. We will see in \ref{sectiondefb}, how one can build a suitable Krein space equipped with the indefinite inner product $[\cdot|\cdot]$ and assign to $b$ a Krein self-adjoint operator on it.

Even though (\ref{chargeproduct}) is not positive definite, it is still possible to give a Hilbert space framework for the Klein-Gordon equation if the potential $V$ is not `too large'. This can be understood in the following way. One introduces the differential operator
\beq\label{adif}
a:=\begin{pmatrix} 0 & 1 \\ \epsilon^2-V^2 &  2 V\end{pmatrix}=\begin{pmatrix} 1 & 0 \\ V &  1\end{pmatrix}\begin{pmatrix} V & 1 \\ \epsilon^2 & V\end{pmatrix}\begin{pmatrix} 1 & 0 \\ -V &  1\end{pmatrix}=\begin{pmatrix} 1 & 0 \\ V &  1\end{pmatrix}b\begin{pmatrix} 1 & 0 \\ -V &  1\end{pmatrix}.
\eeq
In a formal sense, $a$ is symmetric with respect to the sesquilinear form $[ \cdot | \cdot ]_{\rm en}$ (the so-called \textit{energy inner product}), defined as 
\[
[u|v]_{\rm en}:=(u_1|(\epsilon^2-V^2)v_1)+(u_2|v_2)
\]
for suitable elements $u=(u_1,u_2),v=(v_1,v_2)\in L^2 \oplus L^2$. This sesquilinear form is positive if $\epsilon^2-V^2$ is a well-defined \emph{positive} operator. In such a case, one uses $[\cdot|\cdot]_{\rm en}$ to define a Hilbert space and assign to $a$ a self-adjoint operator. One can prove that the Krein self-adjoint operator $b$ is similar to the self-adjoint operator $a$ and therefore has the same spectral properties. Thus, one often prefers to work with $a$ instead of $b$.

The quantity $[ u | u ]_{\rm en}$ is interpreted as energy conserved by the evolution $t\mapsto \e^{\i t a}$ and the violation of positivity, occurring when $\epsilon^2-V^2$ is not positive, is usually called the \textit{Klein paradox} (see \cite{greiner,manogue,fulling} for disambiguation, historical remarks and detailed discussion on the physics of the Klein paradox). In such case it is necessary to work in a Krein space formalism.  Under reasonable restrictions on $V$, one can assign to $a$ a Krein self-adjoint operator and one can prove that it has the same spectrum as the corresponding $b$ operator. The properties of the operator $a$, defined on a suitable Krein space, have been investigated by several authors (there is a particularly vast literature on the positive definite case), see \cite{LNT2} and references therein. 

In this paper, we choose to work with the operator $b$ and the inner product $[\cdot|\cdot]$ only, motivated by the fact that it is closely related to the sesquilinear form used to quantize the space of solutions to the Klein-Gordon equation. Another argument in favor of $b$ is that $(\i\partial_t+b_{(\bx)})$ is easily seen to be prenormally hyperbolic and the analogies to the Dirac Hamiltonian are much more transparent.

The idea of using the operator $b$ for quantization in external potentials dates back to the 1950's \cite{villars}, but we are not aware of a fully rigorous implementation up to date. Many enlightning remarks on quantization in Krein spaces are contained in \cite{schroer}. For applications of Krein spaces in interaction theory, see \cite{gottschalk} and references therein.

\subsection{Operators in Krein spaces}

Let us briefly introduce the notions from Krein space theory needed later on. The standard references are \cite{bognar,langer}. We follow closely the exposition of this subject contained in \cite{gerard} and focus on the class of so called \emph{definitizable Krein self-adjoint} operators, which admits a `smooth function calculus' and `Borel function calculus' with particularly nice properties.  

\begin{definition}\label{defkrein} A Krein space $(\cK, [\cdot|\cdot])$ consists of a Hilbert space $\cK$ with its scalar product $(\cdot|\cdot)_{\cK}$ and an inner product $[ \cdot | \cdot ]$ on $\cK$ (that is a hermitian sesquilinear form), such that $[ \cdot | \cdot ]=(\cdot|g \cdot)_{\cK}$ for some invertible, self-adjoint $g\in B(\cK)$.
\end{definition}

Unless stated otherwise, any topological statements refer to the Hilbert space topology of $\cK$. In the literature, a more general definition of Krein spaces is often used, which requires $\cK$ to be merely a hilbertizable vector space, but this lies away from our case of interest.

Let $A: \cD(A)\to\cK$ be a densely defined operator. The Krein adjoint $A^{\dagger}$ of $A$ in $(\cK, [\cdot|\cdot])$ is defined by
\begin{eqnarray*}
\cD(A^{\dagger}):=\{ u\in\cK : [u|A \cdot ] \mbox{\ is\ continuous\ on\ } \cD(A) \},\\
 \left[ u , A v \right]=[A^{\dagger}u, v] \ \ \ \forall \ u\in\cD(A^{\dagger}), v\in\cD(A).
\end{eqnarray*}
A densely defined operator $A$ is called \emph{Krein self-adjoint}, respectively \emph{Krein unitary} if $A^{\dagger}=A$, resp. $A^{\dagger}A=AA^{\dagger}=1$. It is called \emph{Krein positive} (resp. \emph{Krein negative}) if $[u|Au]\geq 0$ (resp. $\leq0$) for all $u\in\cK$.

\begin{proposition}\label{propran}If $P\in B(\cK)$ is a Krein self-adjoint and Krein positive projection, then $[u|Pu]>0$ for all nonzero $u\in\cK$. Furthermore, $\Ran P$ with scalar product inherited from $\cK$ is a Hilbert space, and its topology coincides with the topology induced by $[u|u]^{1/2}$. 
\end{proposition}

\begin{definition}
A Krein self-adjoint operator $A$ is called \emph{definitizable} if it has non-empty resolvent set and there exists a real polynomial $p(\lambda)$ s.t. $p(A)$ is Krein positive. Such a polynomial is called \emph{definitizing} for $A$. 
\end{definition}

\begin{proposition} Let $A$ be a definitizable operator. Then $\sigma(A)\setminus \R$ consists of finitely many pairs of isolated eigenvalues $\{\lambda_i, \bar\lambda_i \}$.
\end{proposition}

\subsubsection{Smooth functional calculus for definitizable operators}

We quote the adaptation of the function calculus of E.B.~Davies \cite{davies} to definitizable operators in Krein spaces proposed by C.~G\'erard \cite{gerard}, omitting the explicit constructions and proofs. This function calculus is available for classes of smooth functions decreasing fast enough at $\infty$:

For $\rho\in\R$, denote $S^{\rho}(\R)$ the space of functions $f$ such that
\[
\forall_{\alpha\in\N} \ \exists_{C_{\alpha} \geq0}: \ |f^{(\alpha)}(\lambda)|\leq C_{\alpha} \bra\lambda\ket^{\rho-\alpha}\ \ \ ,
\]
equipped with the semi-norms $\|f\|_m:=\sup_{\lambda\in\R,\alpha\leq m}|\bra\lambda\ket^{-\rho+\alpha} f^{(\alpha)}(\lambda)|$. Here, $f^{(\alpha)}$ denotes the derivative of order $\alpha$ of $f$ and $\bra\lambda\ket:=(1+\lambda^2)^{1/2}$. Note that $\cS(\R)\subset S^{\rho}(\R)$ for all $\rho\in\R$.

For $f\in S^{\rho}(\R)$, define
\beq\label{eq:aanalytic}
\tilde f (x+\i y) := \left(\sum^{N}_{r=0}f^{(r)}(x)\frac{(\i y)^r}{r!}\right)\chi\left(\frac{y}{\delta\bra x\ket}\right),
\eeq
where $N$ is some fixed integer, $\delta>0$ and $\chi\in\ccf(\R)$ with $\chi(s)\equiv 1$ for $|s|\leq \half$ and $\chi(s)\equiv 0$ for $|s|\geq 1$. A function defined this way is called an \emph{analytic extension} of $f$. 
It satisfies
\[
\tilde f|_{\R}=f, \ \ \ \ \ \  \left|\frac{\partial\tilde f(z)}{\partial\overline{z}}\right|\leq C \bra \Re z\ket^{\rho-N-1}|\Im z|^N.
\]

\begin{proposition}[\cite{gerard}, B.8]  Let $A$ be a definitizable operator. Let $\rho<-1$, $f\in S^{\rho}(\R)$ and let $\tilde f$ be given by (\ref{eq:aanalytic}). Then for sufficiently high $N$ the integral
\[
f(A):=\frac{1}{2\pi\i}\int_{\C}\frac{\partial\tilde f}{\partial\overline{z}}(A-z)^{-1}\d z\wedge \d \overline{z}.
\] 
is norm convegent in $B(\cK)$ and does not depend on the choice of $\chi$, $\delta$, $N$. The map $S^{\rho}(\R)\ni f\mapsto f(A)\in B(\cK)$ is a homomorphism of algebras and
\beqa
& f(A)^{\dagger}=\overline{f}(A),\\
& \| f(A)\|\leq C_A \|f\|_m, \mbox{ \ for some } m\in\N.
\eeqa
\end{proposition}

It is shown in \cite[B.10]{gerard}, that there is an operator-valued measure $\mu$ such that
\beq
f(A)=\int_{\R}f(t)\d\mu(t) 
\eeq
for each $f\in\ccf(\R)$ with $\supp f\cap\sigma_{\rm cr}(A)=\emptyset$. A construction of such measure $\mu$ is described in \cite{langer}, we will use this particular choice without giving a more explicit characterization. 

\subsubsection{Spectral function for definitizable operators}

A Borel function calculus is also available up to some restrictions \cite{langer,gerard}. There, a crucial role is played by the set of critical points $\sigma_{\rm cr}(A)$, defined as follows.

\begin{definition}
Let $A$ be a definitizable operator. The set
\beq
\sigma_{\rm cr}(A):=\bigcap_p p^{-1}(\{0\})\cap \sigma(A)\cap\R
\eeq
is called the set of critical points of $A$, where the intersection is taken over all definitizing polynomials for $A$.
\end{definition}

\begin{definition}
A finite union of intervals $\cJ\subset\R$ is called \emph{admissible for $A$} if its boundary $\partial\cJ$ contains no critical point of $A$. 
\end{definition}

Let $\cJ\subset\R$ be admissible for $A$. We denote by $\mathfrak{B}_{A}(\cJ)$ the $^*$-algebra of bounded Borel functions on $\cJ$ which are locally constant near $\sigma_{\rm cr}(A)$. 

\begin{theorem}[\cite{gerard}, B.11]\label{borelcalcul} Let $\cJ\subset\R$ be a bounded admissible finite union of intervals for a definitizable operator $A$ and let $g\in\mathfrak{B}_{A}(\cJ)$. Decompose $g=g_0+\sum_i g_i$, where $g_0\in\mathfrak{B}_{A}(\cJ)$ is such that $\supp\, g_0 \cap \sigma_{\rm cr}(A)=\emptyset$ and $g_i\in\ccf(\R)$ ($i=1,\dots,N$; $N<\infty$). Set
\[
g(A):= \sum_{i=1}^N g_i(A) + \int_{\R} g_0(t)\d\mu(t), 
\]
where $g_i(A)$ is defined via smooth functional calculus. Then $g(A)$ is a well-defined operator in $B(\cK)$ and the definition does not depend on the decomposition of $g$. The map
\[
\mathfrak{B}_{A}(\cJ)\ni g \mapsto g(A)\in B(\cK)
\]
is a homomorphism of $^*$-algebras such that $g(A)^{\dagger}=\overline{g}(A)$.
\end{theorem}

We use the Borel functional introduced in Theorem \ref{borelcalcul} to define spectral projections $\one_{\cJ}(A)$, where $\cJ$ is bounded admissible and we recall that $\one_{\cJ}\in\mathfrak{B}_{A}(\cJ)$ denotes the characteristic function of $\cJ$. Equivalently, one could use the construction of spectral projections described in \cite{langer}. To discuss generalizations for larger classes of intervals, one makes the following definition.

\begin{definition}Let $A$ be a definitizable operator. A point $c\in\sigma_{\rm cr}(A)$ is said to be a \emph{regular critical point} of $A$ if $\one_{[c-\varepsilon,c+\varepsilon]}(A)$ converges in the strong operator topology as $\varepsilon \searrow 0$. Otherwise, it is said to be a \emph{singular critical point}. 
We say that a definitizable operator $A$ is \emph{regular at infinity} if $\one_{[-\Lambda,\Lambda]}(A)$ converges in the strong operator topology as $\Lambda\to\infty$.
\end{definition}
Let us note that in the literature, a convention where $\infty$ is by definition in $\sigma_{\rm cr}(A)$ is often employed and one speaks of $\infty$ being a `regular critical point' instead. It is natural to adopt the following notation:

\begin{definition}Let $A$ be a definitizable operator and let $\cJ$ be a finite sum of bounded intervals such that no singular critical points of $A$ intersects $\partial\cJ$. We define 
\beq
\one_{\cJ}(A):=\slim_{\varepsilon\searrow 0} \one_{\cJ(\varepsilon)}(A), \ \ \ \cJ({\varepsilon}):=\cJ\setminus\bigg(\bigcup_{c\in\sigma_{\rm cr}(A)}[c-\varepsilon,c+\varepsilon]\bigg).
\eeq
\end{definition}
\begin{definition}\label{defunbounded}Let $A$ be a definitizable operator, regular at infinity, and let $\cJ$ be a finite sum of intervals such that no singular critical points of $A$ intersects $\partial\cJ$. If $\cJ$ is not bounded, we define 
\beq\label{eqdefreg}
\one_{\cJ}(A):=\slim_{\Lambda\to\infty} \one_{\cJ\cap[-\Lambda,\Lambda]}(A).
\eeq
\end{definition}

\begin{proposition}\label{fcextends}Let $A$ be a definitizable operator and let $\cJ,\cJ'$ be finite sums of bounded intervals such that no singular critical points of $A$ intersects $\partial\cJ$. Let $f\in S^{\rho}(\R)$ with $\rho>-1$ and let $f(A)$ be defined by smooth function calculus. Then:
\begin{enumerate}\setlength{\itemsep}{0pt}
\item $\one_{\cJ}(A)^{\dagger}=\one_{\cJ}(A)$,
\item $\one_{\cJ}(A)\one_{\cJ'}(A)=\one_{\cJ\cap\cJ'}(A)$,
\item if \  $\supp f\cap\overline{\cJ}=\emptyset$ then $f(A)\one_{\cJ}(A)=0$.\label{fcenumber3}
\item if \ $\supp f\subset\overline{\cJ}$ then $f(A)\one_{\cJ}(A)=f(A)$.\label{fcenumber4}
\end{enumerate}
Moreover, if $A$ is regular at infinity, this extends to unbounded $\cJ$ as well.
\end{proposition}
\proof Properties 1.-2. are direct consequences of Theorem \ref{borelcalcul}. To prove properties 3.-4. it suffices to consider $f\in\ccf(\R)$ ($\ccf(\R)$ being dense in $S^{\rho}(\R)$). For such functions the smooth and Borel function calculus concide and using the latter we get
\[
f(A)=(f\cdot\one_{\supp f})(A)=f(A)\one_{\supp f}(A)
\]
and one uses property 2. to get 3.-4. The last assertion follows, as properties 1.-4. are preserved by the strong operator limit (\ref{eqdefreg}).
\qed

\subsubsection{One-parameter groups generated by definitizable  operators }

The following property of definitizable operators which are regular at infinity is essential for our purpose (see \cite{LNT} for a more complete discussion).

\begin{proposition}Let $A$ be definitizable and regular at infinity. Then, it is the generator of a strongly continuous one-parameter group of Krein unitaries $\{ T_{t}\}_{t\in\R}$, i.e.
\[
A x = \lim_{t\to0}\frac{T_{t}x-x}{\i t} \ \ \ \forall x\in\cD(A).
\]
\end{proposition}

Let us now investigate the relation between $T_{t}$ and operators defined by function calculi for $A$. In doing so, one has to take into account that $T_{t}$ contains all the information about eventual complex eigenvalues of $A$, but this not the case for an operator $f(A)$ defined by smooth function calculus. We illustrate this in Proposition \ref{Trelation}. 

First, let us introduce the projection corresponding to the non-real part of the spectrum of $A$. Let $E(\lambda,A)$ denote the Riesz projection relative to an isolated eigenvalue $\lambda\in\sigma(A)$. Define
\beq\label{defonec}
\one^{\C\setminus\R}(A):=\sum_{\lambda\in\sigma(A), \Im\lambda>0}E(\lambda,A)+E(\overline{\lambda},A).
\eeq
A standard result from Krein space theory says that $[u|\one^{\C\setminus\R}(A) u]=0$ for each $u\in\cK$.

\begin{proposition}\label{Trelation} Let $A$ be definitizable and regular at infinity. Denote $\{ T_{t} \}_{t\in\R}$ the one-parameter group it generates. For any $f\in\ccf(\R)$, one has
\beq\label{eq:onepgroups}
\frac{1}{\sqrt{2\pi}}\int_{\R}\d t\,f(t)T_{t} (1-\one^{\C\setminus\R}(A))= (\cF^{-1} f) (A),
\eeq
where $(\cF^{-1} f) (A)$ is defined via smooth functional calculus.
\end{proposition}
\proof By \cite[Theorem 3.12.2]{arendt}, $T_{t}$ is the inverse Laplace transform of the resolvent of $\i A$, i.e. 
\beq\label{eq:laplace1}
T_{t}=\slim_{k\to\infty}\frac{1}{2\pi\i}\int_{-k}^k \e^{(\mu+\i s)t} (A+\i \mu- s)^{-1} \d s
\eeq
for sufficiently large $\mu>0$. By writing the same equality for $T(-t)$ and taking the Krein adjoint, we get also
\beq\label{eq:laplace2}
T_{t}=\slim_{k\to\infty}\frac{(-1)}{2\pi\i}\int_{-k}^k \e^{(-\mu+\i s)t} (A-\i \mu- s)^{-1} \d s.
\eeq
Using (\ref{eq:laplace2}) we get that $\int_{0}^{\infty}\d t\,f(t)T_{t}$ equals 
\begin{eqnarray*}
\frac{(-1)}{2\pi\i}\int_{0}^{\infty}\d t \int_{-\infty}^{\infty}\d s f(t) \e^{(-\mu+\i s)t} (A-\i \mu- s)^{-1}.
\end{eqnarray*}
By the Riesz-Dunford calculus,
\[
(A-w)^{-1}(\one^{\C\setminus\R}(A)-1)=\frac{1}{2\pi\i}\int_{\gamma(w)}(w-z)^{-1}(A-z)^{-1}\d z, 
\]
where $\gamma(w):=\gamma_0(w)\cup\gamma_1\cup\overline{\gamma_1}$, $\gamma_0(w)$ is a circle in $\rs(A)$ which surrounds $w\in\C$ and $\gamma_1$ is a circle in $\rs (A)\cap\{z : \ \Im z > 0 \}$ which surrounds $\sp (A)\cap\{z : \ \Im z > 0 \}$. Hence
\begin{eqnarray*}
\int_{0}^{\infty}\d t\,f(t)T_{t} (1-\one^{\C\setminus\R}(A))\\= \frac{-1}{(2\pi\i)^2}\int_{0}^{\infty}\d t \int_{-\infty}^{\infty}\d s \int_{\gamma(s+\i\mu)}\d z\, f(t) \e^{(-\mu+\i s)t}(s+\i\mu-z)^{-1}(A-z)^{-1}\\=\frac{-1}{(2\pi\i)^2}\int_{0}^{\infty}\d t \int_{-\infty+\i\mu}^{\infty+\i\mu}\d w \int_{\gamma(w)}\d z\, f(t) \e^{\i w t}(w-z)^{-1}(A-z)^{-1}.
\end{eqnarray*}
We claim that the contour $\gamma(w)$ can be replaced by $\eta(\epsilon):=(\R+\i\epsilon)\cup (\R-\i\epsilon)$ (clockwise), where $\epsilon>0$ is arbitrarily small. To this end we have to prove that the respective integral over two half-circles (in the $z$ variables) with center $\i\epsilon$, $-\i\epsilon$ and radius $R$ vanishes as $R\to\infty$. Indeed,  we can use that $\| (A-z)^{-1} \|$ is $O(|\Im z|^{-1})$ for large $|\Im z|$ (as follows from the Hille-Yosida theorem) to show that the integral over $z$ is $O(R^{-1}\ln R)$. We have
\begin{eqnarray*}
\frac{-1}{(2\pi\i)^2}\int_{0}^{\infty}\d t \int_{-\infty+\i\mu}^{\infty+\i\mu}\d w \int_{\eta(\epsilon)}\d z\, f(t) \e^{\i w t}(w-z)^{-1}(A-z)^{-1}\\
=\frac{(-1)}{2\pi\i}\int_{0}^{\infty}\d t \int_{\eta(\epsilon)}\d z\, f(t) \e^{\i z t}(A-z)^{-1}.
\end{eqnarray*}
Analogously, using (\ref{eq:laplace1}) instead of (\ref{eq:laplace2}), one finds
\[
\int_{-\infty}^{0}\d t\,f(t)T_{t} (1-\one^{\C\setminus\R}(A))\\=\frac{(-1)}{2\pi\i}\int_{-\infty}^{0}\d t \int_{\eta(\epsilon)}\d z\, f(t) \e^{\i z t}(A-z)^{-1}.
\]
Thus, denoting $g:=\cF^{-1}f$,
\begin{eqnarray*}
\frac{1}{\sqrt{2\pi}}\int_{\R}\d t\,f(t)T_{t} (1-\one^{\C\setminus\R}(A))=\frac{(-1)}{2\pi\i}\int_{\eta(\epsilon)}g(z)(A-z)^{-1}.
\end{eqnarray*}

On the other hand, to evaluate the RHS of (\ref{eq:onepgroups}), let us note that $g$ is an entire function and it consequently admits an almost analytic extension of the form
\[
\tilde g (z):=g(z)\chi_0(z), \ \ \ \chi_0(x+\i y):=\chi\left(y/(\delta\bra x\ket)\right),
\]
where $\chi$ and $\delta$ are as in \ref{eq:aanalytic}. Therefore,
\begin{eqnarray*}g(A)=\frac{1}{2\pi \i}\int_{\C}\frac{\partial \tilde g}{\partial \bar z}(z)(A-z)^{-1}\d z\wedge \d\bar z
= \lim_{\epsilon\searrow0}\frac{1}{2\pi \i}\int_{C_{\epsilon}}\frac{\partial \tilde g}{\partial \bar z}(z)(A-z)^{-1}\d z\wedge \d\bar z \\
= \lim_{\epsilon\searrow0}\frac{1}{2\pi \i}\int_{\partial C_{\epsilon}}\tilde g(z)(A-z)^{-1}\d z  = \lim_{\epsilon\searrow0}\frac{1}{2\pi \i}\int_{\partial C_{\epsilon}} g(z)\chi_0(z)(A-z)^{-1}\d z,
\end{eqnarray*}

where $C_{\epsilon}:=\supp\chi_0 \cap \{ z: \ |\Im z|>\epsilon\}$. The last integral does not depend on $\epsilon$, hence
\[
g(A)=\frac{1}{2\pi\i}\int_{\bar\eta(\epsilon)}g(z)(A-z)^{-1}=\frac{(-1)}{2\pi\i}\int_{\eta(\epsilon)}g(z)(A-z)^{-1}.
\]
\qed

\subsubsection{Definition and properties of b}\label{sectiondefb}

In this section, we gather results obtained by H.~Langer, B.~Najman and C.~Tretter \cite{LNT}, basing on earlier works (among others) by K.~Veseli\'c \cite{veselic} and P.~Jonas \cite{jonas}.

Following Jonas, they introduce the Hilbert space 
\[
\cK:=\epsilon^{-\half}L^2\oplus\epsilon^{\half}L^2.
\]
More explicitly, $\epsilon^{-1/2}L^2$ is by definition the space $\cD(\epsilon^{\half})$ with scalar product $(\epsilon^{1/2} \cdot | \epsilon^{1/2} \cdot )$ and $\epsilon^{1/2}L^2$ is the completion of $L^2=L^2(\R^d)$ with respect to the norm induced by the scalar product $(\epsilon^{-1/2} \cdot | \epsilon^{-1/2} \cdot )$. Let us note that $\ccf:=\ccf(\R^d,\C^2)$ is dense in $\cK$, as can be easily checked. The indefinite inner product $[ \cdot | \cdot ]$ on $\cK$ is rigorously defined by
\[
[ u | v ]:=(\epsilon^{1/2}u_1|\epsilon^{-1/2}v_2)+(\epsilon^{-1/2}u_2|\epsilon^{1/2}v_1)=(u|g v)_{\cK}
\]
for $u=(u_1,u_2),v=(v_1,v_2)\in\cK$, where $g=\begin{pmatrix} 0 & \epsilon^{-1} \\ \epsilon & 0 \end{pmatrix}$. It follows that $(\cK, [\cdot|\cdot])$ is a Krein space. For $u,v\in\ccf$, we have $[ u | v ]=-\i(u|\sigma^0 v)$ where $\sigma^0=\i\begin{pmatrix} 0 & 1 \\ 1 & 0 \end{pmatrix}$ and $(\cdot|\cdot)$ is here the scalar product in $L^2(\R^d,\C^2)$.

Then, they consider (not necessarily smooth) potentials $V$ satisfying the following assumptions:
\begin{assumption}\label{assumption} $V$ and $\epsilon$ are such that\\
\ \ (i) $\ \cD(\epsilon)\subset\cD(V)$,\\
\ \ (ii) $c=V\epsilon^{-1}$ can be decomposed as $c=c_0+c_1$ with $\|c_0\|<1$ and $c_1$ compact,\\
\ \ (iii) $1\notin \sp_{\rm p}(c^*c)$.
\end{assumption}

The operator $b$ in the Hilbert space $\cK$ is defined by
\[
\cD(b):=\left\{ \begin{pmatrix} v_1 \\ v_2 \end{pmatrix}\in\epsilon^{-\half}L^2\oplus\epsilon^{\half}L^2 : \ v_2\in{L^2}, \ Vv_1+v_2\in \epsilon^{-\half}L^2, \ \epsilon^2 v_1 + V v_2 \in \epsilon^{\half}L^2\right\},\]
\beq \label{defb}
b\begin{pmatrix} v_1 \\ v_2 \end{pmatrix}:=\begin{pmatrix} V v_1 + v_2 \\ \epsilon^2 v_1 + V v_2 \end{pmatrix}.
\eeq

To the differential expression $\epsilon_{(\bx)}^2-V^2(\bx)$ one associates an operator in $\epsilon^{-1}L^2$ given by
\[
\epsilon^2-V^2:=\epsilon (1-c^*c) \epsilon, \ \ \ \ \ \cD(\epsilon^2-V^2):=\{ w\in\epsilon^{-1}L^2: \ (1-c^*c)w\in\epsilon^{-1}L^2\}.
\]

Note that part (iii) of Assumption \ref{mainassumption} is equivalent to  $0\notin\sp_{\rm p}(b)$. This simplifies much the discussion presented later on, but is not an essential assumption and the case $0\in\sp_{\rm p}(b)$ can be treated along the same lines as the analogous problem in the Dirac case. The following theorem summarizes the spectral properties of the operator $b$.

\begin{theorem}[\cite{LNT}]\label{lnttheorem2} Suppose that Assumption \ref{assumption} is satisfied for $c=c_0+c_1$ with $\|c_0\|<1$ and $c_1$ compact, and let $b$ be the operator defined by \ref{defb}. Then:
\begin{innerlist}
\item The operator $b$ is definitizable in the Krein space $(\cK, [\cdot|\cdot])$ and is regular at $\infty$. Consequently, $b$ is the generator of a strongly continuous group of Krein unitaries $\{ T_{t} \}_{t\in\R}$.
\item The essential spectrum $\sigma_{\rm ess}(b)$ is real and $\sp_{\rm ess}(b)\cap (-\alpha,\alpha)=\emptyset$, where $\alpha:=(1-\|c_0\|)\mu$.
\item Assume $\cJ\subset[0,\infty)$ (resp. $\cJ\subset(-\infty,0]$) is admissible for $b$. Then, $\one_{\cJ}(b)$ is Krein positive (resp. Krein negative) iff $\overline{\cJ}\cap \sigma_{\rm cr}(b)=\emptyset$.
\item If $c_1=0$, then $b$ has no complex eigenvalues.
\item If $\epsilon^2-V^2$ is strictly positive, then $\sigma(b)\subset\R$ and $\sigma_{\rm cr}(b)=\emptyset$.
\end{innerlist}
\end{theorem}

In the case $m>0$ and $A_i(\bx)\equiv 0$ for $i=1,\ldots,d$, the operator $\epsilon$ equals $(-\Delta+m^2)^{1/2}$ with domain $W^1_2(\R^d)$. Then, one can give explicit examples of classes of potentials $V$ satisfying the assumptions of Theorem \ref{lnttheorem2}.
\begin{proposition}[\cite{LNT}] Let $d\geq 3$. Parts (i)-(ii) of Assumption \ref{mainassumption} and Assumption \ref{assumption} are satisfied if $\epsilon=(-\Delta+m^2)^{1/2}$ with $m>0$ and $V=V_0+V_1$,
where $V_1\in L^p(\R^d)$ with $d\leq p<\infty$, and one of the following holds:
\begin{enumerate}\setlength{\itemsep}{0pt}
\item $V_0\in L^{\infty}(\R^d)$ with $\|V_0\|_{\infty}<m$;
\item $V_0(\bx)=\gamma/|\bx|$, $\bx\in\R^d\setminus\{0\}$, with $\gamma\in\R$ s.t. $|\gamma|<(d-2)/2$.
\end{enumerate}
\end{proposition}

\subsection{Hadamard distributions}

We introduce the causal propagator $S$ and candidates for positive frequency solutions in analogy to the Dirac case. We allow more freedom in defining the latter in order to treat the overcritical case at once, where it is not clear from the beginning what should be the replacement for $\one_{(0,\infty)}(h)$.

As previously, Assumptions \ref{mainassumption} and \ref{assumption} are assumed, the operator $b$ is defined by (\ref{defb}) and $\{T_t\}_{r\in\R}$ is the one-parameter group generated by $b$. We define multilinear functionals
\beqa\label{eq:defs0}
\bra S,  f\otimes \bar{u} \otimes v \ket:=-\i\int_{\R}f(t)[ u |  T_{t} v ]\d t,\\
\label{eq:defs1}
\bra S^{\cJ}_+, f\otimes \bar{u} \otimes v \ket :=-\i\sqrt{2\pi}\,[ u | (\cF^{-1}f)(b)\one_{\cJ}(b) v ],\\
\label{eq:defs2}
\bra S^{\cJ}_-,  f\otimes \bar{u} \otimes v \ket:=-\i\sqrt{2\pi}\,[ u |  (\cF^{-1}f)(b)( 1-\one_{\cJ}(b) ) v ],
\eeqa
for $f\in\ccf(\R)$, $u,v\in\ccf(\R^d,\C^2)$. Here, $(\cF^{-1}f)(b)$ is defined by smooth function calculus, $\cJ\subset\R$ is a given admissible union of intervals for $b$ and $\one_{\cJ}(b)$ is given by Definition \ref{defunbounded}.

Note that the distributions $S^{\cJ}_{\pm}$ are defined using the smooth function calculus for $b$, which kills any modes with non-real eigenfrequency. This is the reason why they display no exponential behaviour for large times and are consequently tempered in the time variable, as stated in the next proposition.

\begin{proposition}\label{proptempered}The functionals $S^{\cJ}_+, S^{\cJ}_-$ extend to distributions in $\cS'(\R)\hat{\otimes}\cD(\R^{2d},L(\C^2))'$.
\end{proposition}
\proof 
By the Schwartz kernel theorem, (\ref{eq:defs1}) defines uniquely a distribution in the space $\cS'(\R)\hat\otimes\cD(\R^{2d},L(\C^2))'$ if $S^{\cJ}_+:\cS(\R)\otimes\ccf(\R^d,\C^2)\otimes\ccf(\R^d,\C^2)\to\C$ is continuous. By Schwarz inequality and smooth function calculus, we have for some $m\in\N$
\begin{eqnarray*}
{\textstyle\frac{1}{\sqrt{2\pi}}}|\bra S^{\cJ}_+, f \otimes \bar{u} \otimes v\ket|=|[ u |(\cF^{-1}f)(b)\one_{\cJ}(b) v ]|=|( u | g (\cF^{-1}f)(b)\one_{\cJ}(b) v )_{\cK}|
\\ \leq \|u\|_{\cK}  \|(\cF^{-1}f)(b)\| \|g\one_{\cJ}(b)\| \|v \|_{\cK} \leq C_b \|u\|_{\cK}  \|\cF^{-1}f\|_m \|g\one_{\cJ}(b)\| \|v \|_{\cK}.
\end{eqnarray*}
Convergence of $f$ to $0$ in $\cS(\R)$ implies $\cF^{-1} f\to0$ in $\cS(\R)$ and consequently $\| \cF^{-1} f \|_{m}\to 0$. Furthermore, convergence of $u$ (resp. $v$) to $0$ in $\ccf(\R^d,\C^2)$ implies $\left\|u\right\|_{\cK}\to 0$ (resp. $\|v\|_{\cK}$). Indeed, one has the inequality
\[
\|u\|^2_{\cK}=\| \epsilon^{1/2} u_1 \|^2 + \| \epsilon^{-1/2} u_2 \|^2 \leq \| \epsilon^{-3/2} \|^2 \| \epsilon^{2} u_1 \|^2 + \| \epsilon^{-1/2} \|^2 \| u_2 \|^2.
\]
Now, $u_1,u_2\to 0$ in $\ccf(\R^d)$ implies $\epsilon^2 u_1, u_2\to 0$ in $\ccf(\R^d)$ and consequently $\|\epsilon^2 u_1\|$, $\|u_2\| \to 0$.

The reasoning for $S^{\cJ}_-$ is analogous.
\qed

\begin{proposition}\label{bosonproposition} Let $\cJ\subset\R$ be admissible for the operator $b$ (as previously, Assumptions \ref{mainassumption} and \ref{assumption} are assumed and the operator $b$ is defined by (\ref{defb})).
\begin{enumerate}\setlength{\itemsep}{0pt}
\item\label{item1} $S$, $S^{\cJ}_+$ and $S^{\cJ}_-$ are bi-solutions for $\i\partial_t+b_{(\bx)}$,
\item\label{item2} $S$ is the causal propagator for $\i\partial_t+b_{(\bx)}$,
\item\label{item3} If $[\alpha,\infty)\subset\cJ\subset [\alpha',\infty)$ for some $\alpha,\alpha'\in\R$, then $S^{\cJ}_+$ satisfies the static asymptotic spectral condition (cf. Definition \ref{defspectral}).
\end{enumerate}
\end{proposition}

Claims \ref{item1}.-\ref{item2}. are proved as in the spin-0 case.

\textbf{Proof of \ref{item3}., Proposition \ref{bosonproposition}.} \ By Proposition \ref{Trelation}, we have
\[ S=S^{\cJ}_+ + S^{\cJ}_- + S_0,
\]
where $S_0$ restricted to $t=const$ is proportional to the integral kernel of 
\[
T(t)\one^{\C\setminus\R}(b)=\sum_{\lambda\in\sigma(A), \Im\lambda>0}\e^{\i \lambda t}(E(\lambda,A)+E(\overline{\lambda},A)). 
\]
We see that $S_0$ is smooth in the time variable, i.e. $\wf(S_0)\subset \R^{1+2d}\times (\{0\}\times \R^{2d})$. It is also a bi-solution for the pre-normally hyperbolic differential operator $\i\partial_t+b_{(\bx)}$, hence $\wf(S_0)=\emptyset$, that is $S_0$ is smooth.  

By Proposition \ref{proptempered}, both distributions $S^{\cJ}_{\pm}$ are tempered in the time direction, so it remains to prove the assertion on the supports of their Fourier transforms. Note that the distributions, $\cF_0 S^{\cJ}_{+}$, $\cF_0 S^{\cJ}_{-}$ are uniquely determined by their value on simple tensors:
\begin{eqnarray}
\bra \cF_0 S^{\cJ}_{+}, f\otimes \bar{u} \otimes v \ket =-\i\sqrt{2\pi}[ u | f (b)\one_{\cJ}(b) v ],\label{temp1}\\
\bra \cF_0 S^{\cJ}_{-}, f\otimes \bar{u} \otimes v \ket =-\i\sqrt{2\pi}[ u |  f (b)(1-\one_{\cJ}(b))v].\label{temp2}
\end{eqnarray}
By \ref{fcenumber3}. of Proposition \ref{fcextends}, (\ref{temp1}) vanishes for each $f\in\cS(\R)$ with $\supp f \cap \sigma_{\rm cr}(b)=\emptyset$ and $\supp f \cap\overline{\cJ}=\emptyset$. By \ref{fcenumber4}. of Proposition \ref{fcextends}, (\ref{temp2}) vanishes for each $f\in\cS(\R)$ with $\supp f \cap \sigma_{\rm cr}(b)=\emptyset$ and $\supp f\subset \cJ$. Therefore, \[
\supp (\cF_0 S^{\cJ}_{+})\subset (\sigma_{\rm cr}(b)\cup\overline{\cJ})\times\R^{2d}, \ \ \ \supp (\cF_0 S^{\cJ}_{-})\subset (\sigma_{\rm cr}(b)\cup(\R\setminus\cJ))\times\R^{2d},\]
which by boundedness of the set $\sigma_{\rm cr}(b)$ finishes the proof.
\qed

A positivity condition can be formulated in analogy to the Dirac case.
\begin{proposition}\label{bosonicpositivity}Let $\cJ\subset\R$ be admissible for the operator $b$. The following are equivalent:
\begin{enumerate}\setlength{\itemsep}{-1.5pt}
\item $\cJ\subset[0,\infty)$ and $\overline{\cJ}\cap\sigma_{\rm cr}(b)=\emptyset$,
\item $\one_{\cJ}(b)$ is Krein positive,
\item $S_+^{\cJ}$ satisfies the following positivity condition
\beq\label{jpositivity}
-(\tau^* S_+^{\cJ})(\overline{F} \otimes \i F)\geq 0 \ \ \ \forall F\in\ccf(\R^{1,d},\C^2).
\eeq
\end{enumerate}
\end{proposition}
\proof Equivalence (\textit{1}\,$\Leftrightarrow$\textit{2}\,) is part of Theorem \ref{lnttheorem2}. For (\textit{2}\,$\Leftrightarrow$\textit{3}\,), let us remark that (\ref{jpositivity}) is equivalent to
\begin{eqnarray}
\bra -\i \tau^* S^{\cJ}_+, \overline{f\otimes u} \otimes f \otimes u \ket = \sqrt{2\pi} [ u | \overline{(\cF^{-1}f)}(\cF^{-1}f)(b)\one_{\cJ}(b)u ]\nonumber\\
=\sqrt{2\pi}[(\cF^{-1}f)(b)u| \one_{\cJ}(b) (\cF^{-1}f)(b) u]\geq 0 \label{temppositi}
\end{eqnarray}
for all $f\in\ccf(\R^d)$, $u\in\ccf=\ccf(\R^d,\C^2)$. Implication (\textit{2}\,$\Rightarrow$\textit{3}\,) follows. To show that (\ref{temppositi}) implies Krein positivity of $\one_{\cJ}(b)$, fix $f$ s.t. the operator $\cO:=(\cF^{-1}f)(b)$ is invertible, so that (\ref{temppositi}) means $[ \cdot | \one_{\cJ}(b) \cdot ]\geq 0$ on the set $\cO\ccf$. By density of $\ccf$ in $\cK$, $\cO\ccf$ is dense in $\cK$ and the inequality $[ \cdot | \one_{\cJ}(b) \cdot ]\geq 0$ extends to $\cK$.
\qed

As we explain later on in \ref{sectionq}, the positivity condition (\ref{jpositivity}) allows to associate a quasi-free state $\omega_{\cJ}$ to $S_+^{\cJ}$. Therefore, we have found Hadamard states parametrized by admissible sets $\cJ\subset[0,\infty)$. One can easily extend those results to the more general case when $\partial\cJ$ contains no critical singular points of $b$.

We raise now the question of existence of a distinguished Hadamard state. If there are no critical points, the choice $\cJ:=[0,\infty)$ gives rise to the ground state known from other constructions. However, if critical points are present, such choice of $\cJ$ would lead to violation of positivity and $S_+^{\cJ}$ would not define a state in the usual sense. If one insists on preserving positivity, one has to remove all critical points from the interval and consequently $\cJ:=[0,\infty)\setminus\sigma_{\rm cr}(b)$ is the obvious naive choice. If no singular critical points are present, $\one_{\cJ}(b)$ and $S^+_{\cJ}$ are indeed well defined and give rise to a Hadamard state. On the other hand, if there is a singular critical point $c\in[0,\infty)$, $\one_{\cJ}(b)$ is ill-defined and one needs to consider a smaller set $\cJ(\varepsilon):=\cJ\setminus [c-\varepsilon,c+\varepsilon]$. Although $\varepsilon>0$ can be chosen arbitrarily small, none of the sets $\cJ(\varepsilon)$ is distinguished. To clearify what can be a `distinguished state' in this context, we propose the following definition. 
 
\begin{definition}We say that the Hadamard state associated to $S_+^{\cJ}$ is \emph{maximal}, if $\cJ$ is maximal in the directed set $\{ \cJ\subset[0,\infty): \partial\cJ \mbox{\ contains\ no\ singular\ critical\ point\ of\ } b\}$ with relation `$\subset$'. 
\end{definition}

\begin{corollary}\label{finalcoro} Our construction of Hadamard states can be summed up as follows in terms of maximal states.
\begin{enumerate}
\item If $\epsilon^2-V^2$ is positive:\\
There exists a maximal Hadamard state. It corresponds to the choice  $\cJ=[0,\infty)$ and this is precisely the ground state known from other constructions.
\item If $\epsilon^2-V^2$ is not positive and $[0,\infty)$ contains no singular critical point of $b$:\\
There exists a maximal Hadamard state and it corresponds to the choice $\cJ:= [0,\infty) \setminus \sigma_{\rm cr}(b)$.
\item If $\epsilon^2-V^2$ is not positive, and $[0,\infty)$ contains a singular critical point of $b$:\\
There exists no maximal Hadamard state.
\end{enumerate}
\end{corollary}
Sufficient conditions for the second case to hold are given in \cite{gerard}. This includes for instance the case when $A_i\equiv0$, $V\in\ccf(\R^d)\cap L^d(\R^d)$ and $m<\|V\|_{\infty}<\sqrt{2}m$. Unfortunately, we do not know of explicit sufficient conditions for the third case to hold.

\subsection{Quantization}\label{sectionq}

For sake of completeness, we explain the connection between $S_+^{\cJ}$, associated quasi-free states and quantization. Most of the basic facts on bosonic quasi-free states is proved in \cite{araki}, we also use some terminology from \cite{verch} and \cite{derger}. Although our treatment is not standard, as it is based on the operator $b$, it fits into the general framework of bosonic quasi-free states (especially in the mathematical setup originally proposed in \cite{araki}) and recovers known conditions for existence of ground states, see e.g. \cite{broadbridge}.

Let $\cV$ be a real vector space and $\sigma( \cdot,\cdot)$ an antisymmetric form on $\cV$ (not necessarily non-degenerate). Denote $\mathfrak{A}(\cV,\sigma)$ the corresponding Weyl CCR algebra (see e.g. \cite{weylalgebra,derger} for an exact definition), formally generated by elements of the form $W(v)$ for $v\in\cV$, with $W(\cdot)$ satisfying 
\[ W(v)^* = W(-v), \ \ \  W(u)W(v) = \e^{-\i\sigma(u,v)/2} W(u+v), \ \ \ u,v\in\cV. \]

\begin{definition}A state $\omega$ on $\mathfrak{A}(\cV,\sigma)$ is called a \emph{bosonic quasi-free state} if there exists a symmetric form $\mu( \cdot, \cdot)$ on $\cV$ such that
\beq\label{def:qfree}
\omega(W(v))=\e^{-\half\mu(v,v)}, \ \ \ \ \ v\in\cV.
\eeq
\end{definition}

\begin{definition} Let $(\cV, \sigma)$ consists of a real vector space $\cV$ and an antisymmetric form on $\cV$. A real symmetric form $\mu(\cdot,\cdot)$ is said to be dominating for $(\cV, \sigma)$ if
\beq
\mu(u,u)\geq0, \ \ \ |\sigma(u,v)|^2\leq 4 \mu(u,u)\mu(v,v) , \ \ \ u,v\in\cV.
\eeq
\end{definition}

\begin{proposition}\label{defomega} If $\mu$ is a dominating symmetric form for $(\cV, \sigma)$, there exists a unique bosonic quasi-free state $\omega$ on $\mathfrak{A}(\cV,\sigma)$ satisfying (\ref{def:qfree}).
\end{proposition}

The spaces $\epsilon^{\pm1/2}L^2$ have natural complex structures. Denote $\Re \epsilon^{\pm1/2}L^2$ the real Hilbert space consisting of real elements of $\epsilon^{\pm1/2}L^2$ (i.e. all $f\in\epsilon^{\pm1/2}L^2$ s.t. $\bar{f}=f$) and $\Im \epsilon^{\pm1/2}L^2$ the real Hilbert space consisting of imaginary elements of $\epsilon^{\pm1/2}L^2$ (i.e. all $f\in\epsilon^{\pm1/2}L^2$ s.t. $\bar{f}=-f$). Define $\cK_{\R}:=\Re \epsilon^{-1/2}L^2\oplus\Im \epsilon^{1/2}L^2 $ as a real Hilbert space. 

Given $\cJ\subset[0,\infty)$ such that $\partial\cJ$ contains no singular critical point of $b$, we define
\beqa
\sigma(u,v):=-\i[u|v], & u,v\in\cK\label{seconddef1}\\
\mu(u,v):=\sigma(u,{\rm j} v)/2, & u,v\in\cK,\label{seconddef}
\eeqa
where ${\rm j}:=\i(2\cdot\one_{\cJ}(b)-1)$. 
We have
\beqa\label{kahlerproof}
[u|\one_{\cJ}(b)v]=\mu(u,v)+\frac{\i}{2} \sigma(u,v),& u,v\in\cK.
\eeqa
Let us check that $\sigma( \cdot, \cdot)$ is anti-symmetric on $\cK_{\R}$. Indeed,  
\begin{eqnarray*}
\i\sigma( u , v )=(\epsilon^{1/2}u_1|\epsilon^{-1/2}v_2)+(\epsilon^{-1/2}u_2|\epsilon^{1/2}v_1)=
\overline{(\epsilon^{1/2}v_1|\epsilon^{-1/2}u_2)}+\overline{(\epsilon^{-1/2}v_2|\epsilon^{1/2}u_1)}\\
= -(\epsilon^{1/2}v_1|\epsilon^{-1/2}u_2)-(\epsilon^{-1/2}v_2|\epsilon^{1/2}u_1)=-\i\sigma( v , u )
\end{eqnarray*}
for $u,v\in\cK_{\R}$. Together with (\ref{seconddef}-\ref{kahlerproof}) this implies:
\begin{proposition} Let $\cJ\subset\R$ be s.t. $\partial\cJ$ contains no singular critical point of $b$ and let $\mu$, $\sigma$ be given by (\ref{seconddef1}-\ref{seconddef}). Then, $\mu$ is dominating for $(\cK_{\R}, \sigma)$ iff $\cJ\subset[0,\infty)$ and $\cJ\cap\sigma_{\rm cr}(b)=\emptyset$.
\end{proposition}
\proof  We have shown that $\sigma$ is anti-symmetric on $\cK_{\R}$, hence $\sigma(u,u)=0$ for $u\in\cK_{\R}$ and (\ref{kahlerproof}) gives $\mu(u,u)=[u|\one_{\cJ}(b)u]$ for $u\in\cK_{\R}$. For non-negativity of $\mu$, it suffices to read off  conditions for non-negativity of $[\cdot|\one_{\cJ}(b)\cdot]$ given in Proposition \ref{bosonicpositivity}. It follows from (\ref{kahlerproof}) using standard arguments (see e.g. \cite[Lemma 3.3]{araki}) that $\mu$ is dominating.
\qed

We define $\omega_{\cJ}$ to be the state obtained via Proposition \ref{defomega} for the dominating anti-symmetric form $\mu$ for $(\cK_{\R}, \sigma)$.

If $\sigma_{\rm cr}(b)=\emptyset$ and $\cJ=[0,\infty)$, then the sesquilinear form $(\mu_{\C}+\frac{\i}{2}\sigma_{\C})(\cdot,\cdot)=[\cdot|\one_{\cJ}(b)\cdot]$ on $\cK$ is non-degenerate. The (Fock) GNS representation for the state $\omega_{[0,\infty)}$ can be then obtained as follows. The one-particle Hilbert space, denoted $\cZ$, is obtained by complexifying $\cK_{\R}$ using the complex structure ${\rm j}$. The bosonic Fock space $\Gamma(\cZ)$ is obtained in the usual way from the one-particle space via symmetrized tensor products. The operator $b$ on $\cK_{\R}$ promotes to an operator $b_{\cZ}$ on the complex Hilbert space $\cZ$, which is unitarly equivalent to $-\i b{\rm j}=b\,\sgn (b)$ treated as an operator on $\cK$.  

\begin{definition} Let $\R\ni t\mapsto\alpha_t$ be a strongly continuous one-parameter group of automorphisms of a given $C^*$-algebra $\mathfrak{A}$. A state $\omega$ on a $\mathfrak{A}$ is said to be an $\alpha_t$-ground state if $\omega$ is $\alpha_t$-invariant and the generator of $\alpha_t$ in the GNS representation for $\omega$ is a positive operator.
\end{definition}

Under the assumption $\sigma_{\rm cr}(b)=\emptyset$, the state $\omega_{[0,\infty)}$ is invariant under the one-parameter group of automorphisms denoted $t\mapsto\alpha_t$, induced from the symplectic transformation $t\mapsto\e^{\i t b}$ on $(\cK_{\R},\sigma)$. It follows from standard facts on second quantization that positivity of the generator of $\alpha_t$ represented in $\Gamma(\cZ)$ is equivalent to positivity of $b_{\cZ}$. But this is equivalent to positivity of $b\,\sgn (b)$ on $\cK$.

\begin{corollary} $\omega_{\cJ}$ is an $\alpha_t$-ground state if $\epsilon^2-V^2$ is positive and $\cJ=[0,\infty)$.
\end{corollary}

If $\sigma_{\rm cr}(b)\neq\emptyset$, then the sesquilinear form $(\mu_{\C}+\frac{\i}{2}\sigma_{\C})(\cdot,\cdot)=[\cdot|\one_{\cJ}(b)\cdot]$ is degenerate\footnote{We thank J.~Zahn for drawing our attention on this.}, i.e. there exists $u\in\cK$ s.t. $(\mu_{\C}+\frac{\i}{2}\sigma_{\C})(u,u)=0$. The linear span of such vectors is finite dimensional. In the language of \cite{araki}, the corresponding GNS represention is not Fock. In terms used in \cite{wald}, $\omega_{\cJ}$ is not a `regular' state. 

\section{Discussion and outlook}\label{section5}

\setlength{\parindent}{0.6cm}

\hspace{1cm}We have found and characterized basic Hadamard states for the Dirac and Klein-Gordon equation in Minkowski space, coupled to static smooth external potentials. This includes in particular ground states, confirming this way expectations coming from QFT on curved backgrounds and phrased in \cite{marecki}.  

The Hadamard states found in the overcritical Klein-Gordon case are quite peculiar and one may argue they are not very natural, even the `distinguished ones'. In the construction, we needed to remove some points from what is understood as `positive frequency part of the spectrum'. One can also argue that the physical meaning of those states is unclear, as they are constructed in a framework where back-reaction effects are neglected, which cannot be expected to be a meaningful approximation for arbitrarily strong potentials. It is even possible that back-reaction rules out the possibility of creating overcritical potentials at all. Therefore, one should describe this regime in a theory which includes back-reaction effects by treating `semi-classically' the quantum current operator and plugging it into the (Maxwell) equations governing the external electromagnetic field. Then, overcritical Hadamard states and, if they exist, their non-static generalizations, may possibly play the role of unstable solutions of the semi-classical Maxwell equations. 

We did not discuss non-smooth potentials, the formalism of wave front sets not being well adapted to such case. In particular, our proof of \emph{Theorem \ref{basetheorem}} breaks down if the smoothness condition is dropped. It is highly probable that under suitable assumptions, states satisfying the static asymptotic spectral condition (\emph{Definition \ref{defspectral}}) can still be constructed in analogy to the smooth case, even for overcritical potentials. On the other hand, one can expect only some weaker property than the Hadamard condition to hold. 
The analysis of potentials with singularities is motivated both by possible applications in bound state QED \cite{marecki} and simplified models of quantum fields on black hole spacetimes (see e.g. \cite{bachelot} for a superradiant example). 

Naturally, one is also interested in time dependent potentials, especially in view of the applications proposed in the introduction. The lack of translation invariance in the time coordinate makes the problem more difficult, there is also in general no obvious candidate for a distinguished Hadamard state. Still, under some restrictive conditions on the non-static potentials, a second quantized theory is known \cite{rui} and one can ask if any Hadamard states can be associated. Time-zero restrictions of such states should correspond to time-zero restrictions of the states we investigated in the static case.

\setlength{\parindent}{0.0cm}

\appendix
\section{Wave front sets --- basic definitions and properties}
\label{s:appendixa}
\setcounter{equation}{0}
\setcounter{resultcounter}{0}
\renewcommand{\theequation}{A.\arabic{equation}}
\renewcommand{\theresultcounter}{A.\arabic{resultcounter}}

In this Appendix, we gather basic definitions and results from microlocal analysis. The main reference is \cite{hoermander}, for wave front sets of `vectorbundle' distributions we use also \cite{SV01}.

Let $u\in\cD'(\R^p)$. A neighbourhood $\Gamma$ of $k^{\rm o}\in\R^p$ in $\R^p\setminus\{0\}$ is called \emph{conic} if $k\in\Gamma$ implies $\lambda k\in\Gamma$ for all $\lambda>0$. One says that $(x^{\rm o},k^{\rm o})\in\R^p\times(\R^p\setminus \{0\})$ is a \emph{regular directed point} of $u$ if there exists $\varphi\in\ccf(\R^p)$ with $\varphi(x^{\rm o})\neq0$ such that
\[
\forall n\in\N \ \ \exists C_n\in\R \ \mbox{\ s.t. \ } \ |\cF(\varphi u)(k)|\leq C_n (1+|k|)^{-n}
\]
for all $k$ in a conic neighbourhood of $k^{\rm o}$. Here, $\cF(\varphi u)$ denotes the Fourier transform of the compactly supported distribution $\varphi u$.

\begin{definition}
The \emph{wave front set} $\wf(u)$ is defined as the complement in $\R^p\times(\R^p\setminus \{0\})$ of the set of all regular directed points of $u\in\cD'(\R^p)$. 
\end{definition}

Now let $\cE$ be a vector space of dimension $m$. Any distribution $u\in\cD(\R^p,\cE)'$ can be represented as a column of distributions $u_{i}\in\cD'(\R^p)$ with $m$ entries.
\begin{definition} The \emph{wave front set} $\wf(u)$ of $u\in\cD(\R^p,\cE)'$ is defined as
\[
\wf(u)=\bigcup_{i}\wf(u_{i}).\]
\end{definition}

From the definition it is obvious that $\wf(u+v)\subset \wf(u)\cup\wf(v) $ for $u,v\in\cD(\R^p,\cE)'$.

\begin{proposition}Let $u\in\cD(\R^p,\cE)'$. Then, $u\in\cf(\R^p,\cE)$ if and only if $\wf(u)=\emptyset$.
\end{proposition}

\begin{theorem}\label{theoremchar}Let $P:\cf(\R^p,\cE)\to\cf(\R^p,\cE)$ be a differential operator with smooth coefficients and denote $p(x,k)\in\cf(\R^{2p},L(\cE))$ its principal symbol. Then for any $u\in\cD(\R^p,L(\cE))'$ one has 
\[
\wf(P u) \subset \wf(u) \subset  \wf (P u)\cup p^{-1}(\{0\}).
\]
\end{theorem}

{ \small

\newcommand{\etalchar}[1]{$^{#1}$}

}


\begin{thebibliography}{{Man}88}

\bibitem[ABHN11]{arendt} W.~Arendt, C.J.K.~Batty, M.~Hieber and F.~Neubrander, \textsl{Vector-valued Laplace Transforms and Cauchy Problems},
\newblock Springer, 2011.

\bibitem[AG01]{gottschalk}
S.~{Albeverio} and H.~{Gottschalk}, \textsl{ Scattering theory for quantum
  fields with indefinite metric},
\newblock Commun. Math. Phys. \textbf{ 216}, 491--513 (2001).

\bibitem[AHS78]{AHS}
J.~{Avron}, I.~{Herbst} and B.~{Simon}, \textsl{ {Schr\"odinger operators with
  magnetic fields. I: General interactions}},
\newblock Duke Math. J. \textbf{ 45}, 847--883 (1978).

\bibitem[AS71]{araki}
H.~{Araki} and M.~{Shiraishi}, \textsl{ On Quasifree States of the Canonical
  Commutation Relations (I)},
\newblock Publ. Res. Inst. Math. Sci. \textbf{ 7}(1), 105--120 (1971).

\bibitem[{Bac}04]{bachelot}
A.~{Bachelot}, \textsl{ Superradiance and scattering of the charged
  Klein-Gordon field by a step-like electrostatic potential},
\newblock J. Math. Pure Appl. \textbf{ 83}(10), 1179 -- 1239 (2004).

\bibitem[{Bah}10]{bahns}
D.~{Bahns}, \textsl{ {Schwinger Functions in Noncommutative Quantum Field
  Theory}},
\newblock Ann. Henri Poincar{\'e} \textbf{ 11}, 1273--1283 (2010), {0908.4537}.

\bibitem[BF00]{BF00}
R.~{Brunetti} and K.~{Fredenhagen}, \textsl{ {Microlocal Analysis and
  Interacting Quantum Field Theories: Renormalization on Physical
  Backgrounds}},
\newblock Commun. Math. Phys. \textbf{ 208}, 623--661 (2000),
  {arXiv:math-ph/9903028}.

\bibitem[BF09]{BF09}
C.~B{\"a}r and K.~{Fredenhagen}, editors,
\newblock \textsl{ {Quantum Field Theory on Curved Spacetimes}}, volume 786 of
  \textsl{ Lecture Notes in Physics, Berlin Springer Verlag},
\newblock Springer, 2009.

\bibitem[BHR04]{weylalgebra}
E.~{Binz}, R.~{Honegger} and A.~{Rieckers}, \textsl{ {Construction and
  uniqueness of the C*-Weyl algebra over a general pre-symplectic space}},
\newblock J. Math. Phys. \textbf{ 45}, 2885--2907 (2004).

\bibitem[{Bog}74]{bognar}
J.~{Bognar},
\newblock \textsl{ Indefinite Inner Product Spaces},
\newblock Ergebnisse Mathematik und GrenzGeb., Springer, Berlin, 1974.

\bibitem[BR97]{bratteli}
O.~{Bratteli} and D.~W. {Robinson},
\newblock \textsl{ {Operator algebras and quantum statistical mechanics Vol.2:
  Equilibrium states. Statistical mechanics}},
\newblock Springer-Verlag, 1997.

\bibitem[{Bro}83]{broadbridge}
P.~{Broadbridge}, \textsl{ {Existence Theorems for Segal Quantization via
  Spectral Theory in Krein space}},
\newblock Austral. Math. Soc. Ser. B \textbf{ 24}, 439--460 (1983).

\bibitem[BSZ92]{baez}
J.~{Baez}, I.~{Segal} and Z.~{Zhuo},
\newblock \textsl{ Introduction to Algebraic and Constructive Quantum Field
  Theory},
\newblock Princeton University Press, 1992.

\bibitem[{Che}73]{chernoff}
P.~{Chernoff}, \textsl{ Essential self-adjointness of powers of generators of
  hyperbolic equations},
\newblock J. Funct. Anal. \textbf{ 12}(4), 401 -- 414 (1973).

\bibitem[{Dav}95]{davies}
E.~B. {Davies}, \textsl{ The functional calculus},
\newblock J. London Math. Soc \textbf{ 2}, 166--176 (1995).

\bibitem[DG]{derger}
J.~Derezi{\'n}ski and C.~G{\'e}rard,
\newblock Mathematics of Quantization and Quantum Fields,
\newblock in preparation.

\bibitem[DG10]{positive}
J.~Derezi{\'n}ski and C.~G{\'e}rard, \textsl{ Positive Energy Quantization of
  Linear Dynamics},
\newblock Banach Center Publ. \textbf{ 89}, 75--104 (2010).

\bibitem[Dim82]{dimock}
J.~Dimock, \textsl{ Dirac quantum fields on a manifold},
\newblock Trans. AMS \textbf{ 269}, 133--147 (1982).

\bibitem[DM75]{dosch}
H.~G. {Dosch} and V.~F. M{\"u}ller, \textsl{ Renormalization of Quantum
  Electrodynamics in an Arbitrarily Strong Time Independent External Field},
\newblock Fortschr. Phys. \textbf{ 23}(11-12), 661--689 (1975).

\bibitem[DMP09]{DMP}
C.~{Dappiaggi}, V.~{Moretti} and N.~Pinamonti, \textsl{ Distinguished quantum states in a class of cosmological spacetimes and their Hadamard property },
\newblock J. Math. Phys. \textbf{ 50}, 062304--062304-38 (2009).

\bibitem[FGS91]{fradkin}
E.~S. {Fradkin}, D.~M. {Gitman} and S.~M. {Shvartsman},
\newblock \textsl{ Quantum Electrodynamics With Unstable Vacuum},
\newblock Springer-Verlag, 1991.

\bibitem[{Ful}89]{fulling}
S.~A. {Fulling},
\newblock \textsl{ {Aspects of Quantum Field Theory in Curved Spacetime}},
\newblock Cambridge University Press, 1989.

\bibitem[FV58]{villars}
H.~{Feshbach} and F.~{Villars}, \textsl{ {Elementary Relativistic Wave
  Mechanics of Spin 0 and Spin 1/2 Particles}},
\newblock Rev. Mod. Phys. \textbf{ 30}, 24--45 (1958).

\bibitem[G{\'e}r11]{gerard}
C.~G{\'e}rard, \textsl{ {Scattering theory for Klein-Gordon equations with
  non-positive energy}},
\newblock Ann. Henri Poincar{\'e} (2011), {1424-0637}.

\bibitem[GMR85]{greiner}
W.~{Greiner}, B.~M{\"u}ller and J.~{Rafelski},
\newblock \textsl{ {Quantum Electrodynamics of Strong Fields}}, volume 440 of
  \textsl{ Lecture Notes in Physics},
\newblock Springer, 1985.

\bibitem[{Hac}10]{hack}
T.-P. {Hack},
\newblock \textsl{ {On the Backreaction of Scalar and Spinor Quantum Fields in
  Curved Spacetimes --- From the Basic Foundations to Cosmological
  Applications}},
\newblock PhD thesis, DESY-THESIS-2010-042, 2010.

\bibitem[{Hol}01]{hollands}
S.~{Hollands}, \textsl{ {The Hadamard Condition for Dirac Fields and Adiabatic
  States on Robertson-Walker Spacetimes}},
\newblock Commun. Math. Phys. \textbf{ 216}, 635--661 (2001),
  {arXiv:gr-qc/9906076}.

\bibitem[H{\"o}r83]{hoermander}
L.~H{\"o}rmander,
\newblock \textsl{ The Analysis of Linear Partial Differential Operators I.
  Distribution Theory and Fourier Analysis},
\newblock Springer-Verlag, 1983.

\bibitem[HW05]{hw05}
S.~{Hollands} and R.~M. {Wald}, \textsl{ {Conservation of the Stress Tensor in
  Perturbative Interacting Quantum Field Theory in Curved Spacetimes}},
\newblock Rev. Math. Phys. \textbf{ 17}, 227--311 (2005),
  {arXiv:gr-qc/0404074}.

\bibitem[{Jin}00]{jin}
W.~M. {Jin}, \textsl{ {Quantization of Dirac fields in static spacetime}},
\newblock Classical Quant. Grav. \textbf{ 17}, 2949--2964 (2000),
  {arXiv:gr-qc/0009010}.

\bibitem[{Jon}88]{jonas}
P.~{Jonas}, \textsl{ On a class of selfadjoint operators in Krein space and
  their compact perturbations},
\newblock Integr. Equat. Operat. Theor. \textbf{ 11}, 351--384 (1988),
\newblock 10.1007/BF01202078.

\bibitem[KES{\etalchar{+}}91]{meanfield}
Y.~{Kluger}, J.~M. {Eisenberg}, B.~{Svetitsky}, F.~{Cooper} and E.~{Mottola},
  \textsl{ {Pair production in a strong electric field}},
\newblock Phys. Rev. Lett. \textbf{ 67}, 2427--2430 (1991).

\bibitem[{Kra}00]{kratz}
K.~{Kratzert}, \textsl{ Singularity structure of the two point function of the
  free Dirac field on a globally hyperbolic spacetime},
\newblock Annalen Phys. \textbf{ 9}, 475--498 (2000).

\bibitem[KW91]{kaywald}
B.~S. {Kay} and R.~M. {Wald}, \textsl{ {Theorems on the uniqueness and thermal
  properties of stationary, nonsingular, quasifree states on spacetimes with a
  bifurcate killing horizon}},
\newblock Phys. Rep. \textbf{ 207}, 49--136 (1991).

\bibitem[{Lan}82]{langer}
H.~{Langer},
\newblock Spectral functions of definitizable operators in Krein spaces,
\newblock in \textsl{ Functional Analysis}, edited by D.~Butkovic, H.~Kraljevic
  and S.~Kurepa, volume 948 of \textsl{ Lecture Notes in Mathematics}, pages
  1--46, Springer Berlin / Heidelberg, 1982,
\newblock 10.1007/BFb0069840.

\bibitem[LNT06]{LNT2}
H.~{Langer}, B.~{Najman} and C.~{Tretter}, \textsl{ {Spectral Theory of the
  Klein-Gordon Equation in Pontryagin Spaces}},
\newblock Commun. Math. Phys. \textbf{ 267}, 159--180 (2006).

\bibitem[LNT08]{LNT}
H.~{Langer}, B.~{Najman} and C.~{Tretter}, \textsl{ {Spectral Theory of the
  Klein-Gordon Equation in Krein Spaces}},
\newblock Proc. Edinburgh Math. Soc. \textbf{ 51}(03), 711--750 (2008).

\bibitem[{Man}88]{manogue}
C.~A. {Manogue}, \textsl{ {The Klein paradox and superradiance}},
\newblock Ann. Phys. \textbf{ 181}, 261--283 (1988).

\bibitem[Mar03]{marecki}
P.~Marecki,
\newblock \textsl{ {Quantum electrodynamics on background external fields}},
\newblock PhD thesis, DESY-THESIS-2004-002, 2003.

\bibitem[{Mor}03]{moretti}
V.~{Moretti}, \textsl{ {Comments on the Stress-Energy Tensor Operator in Curved
  Spacetime}},
\newblock Commun. Math. Phys. \textbf{ 232}, 189--221 (2003),
  {arXiv:gr-qc/0109048}.

\bibitem[MPS98]{mohr}
P.~J. {Mohr}, G.~{Plunien} and G.~{Soff}, \textsl{ {QED corrections in heavy
  atoms}},
\newblock Phys. Rep. \textbf{ 293}, 227--369 (1998).

\bibitem[MS05]{shubin}
V.~{Maz'ya} and M.~{Shubin}, \textsl{ {Discreteness of spectrum and positivity
  criteria for Schr{\"o}dinger operators}},
\newblock Ann. Math. \textbf{ 162}, 919 -- 942 (2005), {arXiv:math/0305278}.

\bibitem[M{\"u}h11]{muehlhoff}
R.~M{\"u}hlhoff, \textsl{ {Cauchy problem and Green's functions for first order
  differential operators and algebraic quantization}},
\newblock J. Math. Phys. \textbf{ 52}(2), 022303--+ (2011), {1001.4091}.

\bibitem[{Rad}96]{radzikowski}
M.~J. {Radzikowski}, \textsl{ {Micro-local approach to the Hadamard condition
  in quantum field theory on curved space-time}},
\newblock Commun. Math. Phys. \textbf{ 179}, 529--553 (1996).

\bibitem[{Rin}01]{ringwald}
A.~{Ringwald}, \textsl{ {Fundamental physics at an X-ray free electron laser}},
\newblock Phys. Lett. B \textbf{ 510}, 107 (2001), {arXiv:hep-ph/0112254}.

\bibitem[Rui77]{rui}
S.~N.~M. Ruijsenaars, \textsl{ Charged particles in external fields II. The
  quantized Dirac and Klein-Gordon theories},
\newblock Commun. Math. Phys. \textbf{ 52}, 267--294 (1977),
\newblock 10.1007/BF01609487.

\bibitem[RVX10]{ruffini}
R.~{Ruffini}, G.~{Vereshchagin} and S.-S. {Xue}, \textsl{ {Electron-positron
  pairs in physics and astrophysics: From heavy nuclei to black holes}},
\newblock Phys. Rep. \textbf{ 487}, 1--140 (2010), {0910.0974}.

\bibitem[{San}10]{sanders}
K.~{Sanders}, \textsl{ The locally covariant Dirac field},
\newblock Rev. Math. Phys. \textbf{ 22}, 381--430 (2010).

\bibitem[{Sch}95]{scharf}
G.~{Scharf},
\newblock \textsl{ Finite Quantum Electrodynamics: The Causal Approach; 2nd
  ed.},
\newblock Texts and monographs in physics, Springer, Berlin, 1995.

\bibitem[Shi91]{shi}
I.~Shigekawa, \textsl{ Spectral properties of Schr{\"o}dinger operators with
  magnetic fields for a spin {1/2} particle},
\newblock J. Funct. Anal. \textbf{ 101}(2), 255 -- 285 (1991).

\bibitem[SS70]{schroer}
B.~{Schroer} and J.~A. {Swieca}, \textsl{ {Indefinite Metric and Stationary
  External Interactions of Quantized Fields}},
\newblock Phys. Rev. \textbf{ 2}, 2938--2943 (1970).

\bibitem[SV00]{SV00}
H.~{Sahlmann} and R.~{Verch}, \textsl{ {Passivity and Microlocal Spectrum
  Condition}},
\newblock Commun. Math. Phys. \textbf{ 214}, 705--731 (2000),
  {arXiv:math-ph/0002021}.

\bibitem[SV01]{SV01}
H.~{Sahlmann} and R.~{Verch}, \textsl{ {Microlocal Spectrum Condition and
  Hadamard Form for Vector-Valued Quantum Fields In Curved Spacetime}},
\newblock Rev. Math. Phys. \textbf{ 13}, 1203--1246 (2001),
  {arXiv:math-ph/0008029}.

\bibitem[SVW02]{reeh}
A.~{Strohmaier}, R.~{Verch} and M.~{Wollenberg}, \textsl{ {Microlocal analysis
  of quantum fields on curved space-times: Analytic wave front sets and
  Reeh-Schlieder theorems}},
\newblock J. Math. Phys. \textbf{ 43}, 5514--5530 (2002),
  {arXiv:math-ph/0202003}.

\bibitem[{Tha}92]{thaller}
B.~{Thaller},
\newblock \textsl{ The Dirac Equation},
\newblock Texts and monographs in physics, Springer, Berlin, 1992.

\bibitem[{Ver}97]{verch}
R.~{Verch}, \textsl{ Continuity of symplectically adjoint maps and the
  algebraic structure of Hadamard vacuum representations for quantum fields on
  curved spacetime},
\newblock Rev. Math. Phys. \textbf{ 9}, 635--674 (1997).

\bibitem[Ves70]{veselic}
K.~Veseli{\'c}, \textsl{ {A spectral theory for the Klein-Gordon equation with
  an external electrostatic potential}},
\newblock Nucl. Phys. \textbf{ 147}, 215--224 (1970).

\bibitem[{Wal}77]{wald77}
R.~M. {Wald}, \textsl{ {The back reaction effect in particle creation in curved
  spacetime}},
\newblock Commun. Math. Phys. \textbf{ 54}, 1--19 (1977).

\bibitem[{Wal}94]{wald}
R.~M. {Wald},
\newblock \textsl{ Quantum Field Theory In Curved Spacetime and Black Hole
  Thermodynamics},
\newblock University of Chicago Press, 1994.

\bibitem[{Zah}11]{zahn}
J.~{Zahn}, \textsl{ {Divergences in Quantum Field Theory on the Noncommutative
  Two-Dimensional Minkowski Space with Grosse--Wulkenhaar Potential}},
\newblock Ann. Henri Poincar{\'e} \textbf{ 12}, 777--804 (2011), {1005.0541}.

\end{thebibliography}
\end{document}